\documentclass[12pt,a4paper,english,eqsecnum]{article}
\usepackage[T1]{fontenc}
\usepackage[utf8]{inputenc}
\usepackage{amsmath}
\usepackage{amssymb}
\usepackage{graphicx}
\usepackage{esint}

\makeatletter

\providecommand{\tabularnewline}{\\}

\@ifundefined{definecolor}
 {\usepackage{color}}{}
\usepackage{a4wide}\usepackage{psfrag}
\renewcommand{\theequation}{\hbox{\normalsize\arabic{section}.\arabic{equation}}}
\@addtoreset{equation}{section}
\renewcommand{\thefigure}{\hbox{\normalsize\arabic{section}.\arabic{figure}}}
\@addtoreset{figure}{section}
\renewcommand{\thetable}{\hbox{\normalsize\arabic{section}.\arabic{table}}}
\@addtoreset{table}{section}

\newcommand{\ket}[1]{{\left|#1\right\rangle}}\newcommand{\bra}[1]{{\left\langle #1\right|}}

\usepackage{ifpdf}\ifpdf\usepackage{epstopdf}\usepackage[pdftex,ps2pdf,dvips,colorlinks,urlcolor=blue,citecolor=blue,linkcolor=blue]{hyperref}\else
\usepackage[hypertex,ps2pdf,dvips,colorlinks,urlcolor=blue,citecolor=blue,linkcolor=blue]{hyperref}\fi\pdfadjustspacing=1

\usepackage{babel}

\makeatother

\usepackage{babel}
\begin{document}

\title{Spectral expansion for finite temperature two-point functions and
clustering}

\author{I.M. Szécsényi$^{1}$ and G. Takács$^{2,3}$\\
 ~\\
 $^{1}$Department of Theoretical Physics, Eötvös University\\
 1117 Budapest, Pázmány Péter sétány 1/A, Hungary\\
 ~\\
 $^{2}$Department of Theoretical Physics, \\
 Budapest University of Technology and Economics\\
1111 Budapest, Budafoki út 8, Hungary\\
 ~\\
 $^{3}$MTA-BME \textquotedbl{}Momentum\textquotedbl{} Statistical
Field Theory Research Group\\
1111 Budapest, Budafoki út 8, Hungary}

\date{1st October 2012}
\maketitle
\begin{abstract}
Recently, the spectral expansion of finite temperature two-point functions
in integrable quantum field theories was constructed using a finite
volume regularization technique and the application of multidimensional
residues. In the present work, the original calculation is revisited.
By clarifying some details in the residue evaluations, we find and
correct some inaccuracies of the previous result. The final result
for contributions involving no more than two particles in the intermediate
states is presented. The result is verified by proving a symmetry
property which follows from the general structure of the spectral
expansion, and also by numerical comparison to the discrete finite
volume spectral sum. A further consistency check is performed by showing
that the expansion satisfies the cluster property up to the order
of the evaluation. 
\end{abstract}

\section{Introduction}

Correlation functions play a central role in the formulation of many-body
quantum systems. Integrable models presents a unique opportunity to
study strongly correlated quantum systems in situations where conventional
methods break down. Recent experimental advances resulted in renewed
interest in integrable models, since it is now possible to realize
certain models with the help of optical and magnetic traps \cite{low-D-trapped,2003PhRvL..91y0402M,2003cond.mat.12003L,vanDruten-YangYang}
or in low-dimensional magnets \cite{Coldea:2010,Tennant:2012aa}.

In a recent paper \cite{Pozsgay:2010cr} finite temperature (i.e.
thermal) two-point correlation functions were constructed using the
exact form factors in 1+1 dimensional integrable models. In an integrable
quantum field theory, the basic object is the factorized S-matrix
\cite{zam-zam,Mussardo:1992uc}. The matrix elements of the local
operators (form factors) satisfy a certain set of equations (the form
factor bootstrap equations) which follow from general field theoretical
arguments supplemented with the special analytic properties of the
S-matrix \cite{Karowski:1978vz,Smirnov:1992vz,zam_Lee_Yang,Delfino:1996nf}.
Solving these equations gives the form factor functions, which can
then be used to construct correlation functions by expanding in the
basis formed by the infinite volume asymptotic scattering states.

The form factor expansion of zero-temperature correlations in integrable
QFT is very well understood. In general, the series has very good
convergence properties in massive models and can be evaluated numerically
to any desired precision \cite{zam_Lee_Yang,ising_ff1}. However,
the problem of thermal correlation functions is much more complicated
and has been the subject of active research in the last two decades
\cite{Leclair:1996bf,Leclair:1999ys,Saleur:1999hq,Lukyanov:2000jp,Delfino:2001sz,Mussardo:2001iv,Konik:2001gf,Essler:2004ht,Altshuler:2005ty}.
The form factor construction of the spectral series is plagued with
problems due to the presence of disconnected terms in the expansion,
which lead to formally divergent expressions. Following Balog it can
be shown that the divergent parts cancel with contributions from the
partition function \cite{Balog:1992gf}. Nevertheless, it is a very
non-trivial task to obtain the correct finite answer. Leclair and
Mussardo conjectured an answer for the spectral expansion for one-point
and two-point functions in terms of form factors dressed by appropriate
occupation number factors containing the pseudo-energy function from
the thermodynamical Bethe Ansatz \cite{Leclair:1999ys}. Their proposal
for the two-point function was questioned by Saleur \cite{Saleur:1999hq};
on the other hand, in the same paper he also gave a proof of the Leclair-Mussardo
formula for one-point functions provided the operator considered is
the density of some local conserved charge. By comparison to an alternative
proposal \cite{Delfino:2001sz}, it was also shown that the results
obtained by naive regularization are ambiguous \cite{Mussardo:2001iv}.

The idea behind our approach is to use a finite volume setting to
regularize the divergences. In \cite{Pozsgay:2007kn,Pozsgay:2007gx}
this was applied to one-point functions giving a confirmation of the
Leclair-Mussardo formula to third order; later a derivation to all
orders was also obtained \cite{Pozsgay:2010xd}. The crucial point
is that finite volume is not an ad hoc, but a physical regulator (since
physically realizable systems are always of finite size), therefore
one is virtually guaranteed to obtain the correct result when taking
the infinite volume limit. The natural small parameter for the finite
temperature expansion is the Boltzmann-factor $e^{-m/T}$ where $m$
is the mass gap (which is assumed to be nonzero). The result is an
integral series, where the $N$th term represents $N$-particle processes
over the Fock-vacuum. The contributions with a low number of particles
can be interpreted as disconnected terms of matrix elements calculated
in a thermal state with a large number of particles \cite{Altshuler:2005ty,Pozsgay:2010xd}.
In this sense the approach is similar to the one used in algebraic
Bethe Ansatz \cite{korepin-LL1,iz-kor-resh}.

Besides correlation functions, the finite volume regularization can
also be applied to numerous other problems. The finite volume form
factor approach was extended to boundary operators as well \cite{Kormos:2007qx},
which was used to compute finite temperature one-point functions of
boundary operators \cite{Takacs:2008ec}. Another application of the
bulk finite volume form factors is the construction of one-point functions
of bulk operators on a finite interval \cite{Kormos:2010ae}. It was
also used \cite{Takacs:2009fu} to construct the form factor perturbation
expansion in non-integrable field theories (originally proposed by
Delfino et al. \cite{Delfino:1996xp}) beyond the leading order. It
also turned out that this approach can be applied to quenches in field
theory \cite{Kormos:2010ae,Schuricht:2012aa,Calabrese:2012aa}.

Regarding thermal two-point functions, the finite volume regularization
method was first applied by Essler and Konik \cite{Essler:2007jp,Essler:2009zz};
however, their methods do not have any obvious extension to higher
order. Despite this shortcoming, their results are very useful as
shown by their relevance to inelastic neutron scattering experiments
\cite{Tennant:2012aa}. An independent early calculation of the one-particle--one-particle
contribution can also be found in \cite{Pozsgay:2009pv}.

In \cite{Pozsgay:2010cr} we developed a systematic method to compute
the finite temperature form factor expansion to arbitrary orders.
It turns out that the machinery of multidimensional residues provides
an appropriate formalism to evaluate higher order corrections systematically,
and this was demonstrated to all orders which involve only intermediate
states with at most two particles. To verify the result, we applied
two consistency checks. The first of them was that the correlator
should have a finite limit when the volume is taken to infinity, therefore
all terms containing positive powers of the volume had to cancel,
which was indeed true. The second one took into account that for some
contributions there are two independent ways to arrive at the answer,
and agreement between them also provides validating evidence. However,
for the term $D_{22}$, which contains the contributions when both
intermediate states involved in the spectral sum contain two particles,
the second one is not available and the first is insufficient to check
the structure of the result in detail.

Therefore we decided to provide a numerical evidence, especially since
the analytic manipulations themselves are rather tedious and complicated,
with many possible sources of mistakes. As it turned out, the result
for $D_{22}$ reported in \cite{Pozsgay:2010cr} is unfortunately
incorrect. By further investigation, it turned out that some fine
details of the residue calculation needed to be carried out more carefully. 

Here we report the correct version of the computation and its result,
and present the final formula for the finite temperature two-point
function including all contributions with at most two-particle intermediate
states. To be confident in our results, we perform several checks.
First we check that the result for $D_{22}$ satisfies a particular
symmetry property following from the general form of the spectral
expansion. Then we apply a detailed numerical verification of our
analytic manipulations, and also verify that the final result satisfies
the physically required cluster property.

The layout of the paper is as follows. Section \ref{sec:Finite-volume-regularization}
introduces the thermal two-point function and the idea of finite volume
regularization. In section \ref{sec:The-spectral-expansion}, the
methods to evaluate the resulting spectral expansion is presented.
We re-derive the results of \cite{Pozsgay:2010cr}, including the
correct form of the term $D_{22}$ and present the full formula of
the two-point function including all two-particle contributions. The
numerical verification of $D_{22}$ is performed in Section \ref{sec:Numerical-verification-of},
together with a similar verification for $D_{12}$ in order to establish
a benchmark point for numerical accuracy. In Section \ref{sec:The-cluster-property}
we prove that the resulting expansion satisfies the cluster property,
and in Section \ref{sec:Conclusions} the conclusions are presented.
There are also three appendices: Appendix \ref{sec:Residue-evaluations}
contains the mathematical formulas used for evaluating the residue
contributions, while Appendix \ref{sec:Pole-terms-for} contains the
end results of the residue evaluations, which are also necessary for
the numerical comparison. The proof of the symmetry property of the
$D_{22}$ contribution is given in Appendix \ref{sec:Symmetry-of-D22}.

\section{Finite volume regularization\label{sec:Finite-volume-regularization}}

\subsection{The thermal two-point function\label{sub:The-thermal-two-point}}

A field theory with finite temperature $T$ can be defined using a
compact Euclidean (Matsubara) time $t$: 
\begin{equation}
t\equiv t+R\quad\mathrm{where}\quad R=1/T\label{eq:temperature}
\end{equation}
We are interested in the two-point function in $1+1$ dimensional
field theories: 
\begin{equation}
\langle\mathcal{O}_{1}(x,t)\mathcal{O}_{2}(0)\rangle^{R}=\frac{\mathrm{Tr}\left(\mathrm{e}^{-RH}\mathcal{O}_{1}(x,t)\mathcal{O}_{2}(0)\right)}{\mathrm{Tr}\left(\mathrm{e}^{-RH}\right)}\label{finiteTcorr}
\end{equation}
A naive spectral sum leads to an ill-defined expression due to the
presence of disconnected contributions (cf. e.g. the discussion in
\cite{Pozsgay:2010cr}). However, one can put the system in a finite
spatial volume $L$ with periodic boundary conditions 
\begin{equation}
x\equiv x+L\label{PBC}
\end{equation}
so that 
\begin{equation}
\langle\mathcal{O}_{1}(x,t)\mathcal{O}_{2}(0)\rangle_{L}^{R}=\frac{\text{Tr}_{L}\left(\mathrm{e}^{-RH_{L}}\mathcal{O}_{1}(x,t)\mathcal{O}_{2}(0)\right)}{\text{Tr}_{L}\left(\mathrm{e}^{-RH_{L}}\right)}\label{eq:finvolfiniteTcorr}
\end{equation}
where $\mathrm{Tr}_{L}$ denotes the trace over the finite-volume
states, $H_{L}$ is the Hamiltonian in volume $L$. This expression
can be expanded inserting two complete sets of states 
\begin{equation}
\text{Tr}_{L}\left(\mathrm{e}^{-RH_{L}}\mathcal{O}_{1}(x,t)\mathcal{O}_{2}(0)\right)=\sum_{m,n}\mathrm{e}^{-RE_{n}(L)}\langle n|\mathcal{O}_{1}(x,t)|m\rangle_{L}\langle m|\mathcal{O}_{2}(0)|n\rangle_{L}\label{eq:2ptexp}
\end{equation}
where the matrix elements of local operators are also taken in the
finite volume system. To evaluate it, we need an expression for form
factors in finite volume.

\subsection{The form factor bootstrap\label{sub:The-form-factor}}

In a $1+1$ dimensional field theory, the energy and the momentum
of an on-shell particle is parametrized by the rapidity variable as
$E=m\cosh\theta$ and $p=m\sinh\theta$. For the sake of simplicity
let us suppose that the spectrum of the model consists of a single
particle mass $m$. Incoming and outgoing asymptotic states are defined
as: 
\[
|\theta_{1},\dots,\theta_{n}\rangle=\begin{cases}
|\theta_{1},\dots,\theta_{n}\rangle^{in} & :\;\theta_{1}>\theta_{2}>\dots>\theta_{n}\\
|\theta_{1},\dots,\theta_{n}\rangle^{out} & :\;\theta_{1}<\theta_{2}<\dots<\theta_{n}
\end{cases}
\]
Integrability leads to factorized scattering, which can be summarized
by the relation 
\[
|\theta_{1},\dots,\theta_{k},\theta_{k+1},\dots,\theta_{n}\rangle=S(\theta_{k}-\theta_{k+1})|\theta_{1},\dots,\theta_{k+1},\theta_{k},\dots,\theta_{n}\rangle
\]
where $S$ denotes the two-particle amplitude; from this any multi-particle
scattering process can be obtained by reordering the particles. States
are normalized as: 
\begin{equation}
\langle\theta'|\theta\rangle=2\pi\delta(\theta'-\theta)
\end{equation}
The form factors of a local operator $\mathcal{O}(t,x)$ are defined
as 
\begin{equation}
F_{mn}^{\mathcal{O}}(\theta_{1}',\dots,\theta_{m}'|\theta_{1},\dots,\theta_{n})_{=}\langle\theta_{1}',\dots,\theta_{m}'\vert\mathcal{O}(0,0)\vert\theta_{1},\dots,\theta_{n}\rangle\label{eq:genff}
\end{equation}
With the help of the crossing relations 
\begin{eqnarray}
 &  & F_{mn}^{\mathcal{O}}(\theta_{1}',\dots,\theta_{m}'|\theta_{1},\dots,\theta_{n})=F_{m-1n+1}^{\mathcal{O}}(\theta_{1}',\dots,\theta_{m-1}'|\theta_{m}'+i\pi,\theta_{1},\dots,\theta_{n})\label{eq:crossing}\\
 &  & +\sum_{k=1}^{n}2\pi\delta(\theta_{m}'-\theta_{k})\prod_{l=1}^{k-1}S(\theta_{l}-\theta_{k})F_{m-1n-1}^{\mathcal{O}}(\theta_{1}',\dots,\theta_{m-1}'|\theta_{1},\dots,\theta_{k-1},\theta_{k+1}\dots,\theta_{n})\nonumber 
\end{eqnarray}
all form factors can be expressed in terms of the elementary form
factors 
\[
F_{n}^{\mathcal{O}}(\theta_{1},\dots,\theta_{n})=\langle0\vert\mathcal{O}(0,0)\vert\theta_{1},\dots,\theta_{n}\rangle
\]
which satisfy the form factor bootstrap equations \cite{Karowski:1978vz,Kirillov:1987jp,Smirnov:1992vz}
\begin{eqnarray}
\mbox{Lorentz symmetry: } &  & F_{n}^{\mathcal{O}}(\theta_{1}+\Lambda,\theta_{2}+\Lambda,\dots,\theta_{n}+\Lambda)=\label{eq:shiftaxiom}\\
 &  & \qquad\exp\left(s_{\mathcal{O}}\Lambda\right)F_{n}^{\mathcal{O}}(\theta_{1},\theta_{2},\dots,\theta_{n})\nonumber \\
\mbox{Exchange:} &  & F_{n}^{\mathcal{O}}(\theta_{1},\dots,\theta_{k},\theta_{k+1},\dots,\theta_{n})=\label{eq:exchangeaxiom}\\
 &  & \qquad S(\theta_{k}-\theta_{k+1})F_{n}^{\mathcal{O}}(\theta_{1},\dots,\theta_{k+1},\theta_{k},\dots,\theta_{n})\nonumber \\
\mbox{Cyclic property:} &  & F_{n}^{\mathcal{O}}(\theta_{1}+2i\pi,\theta_{2},\dots,\theta_{n})=F_{n}^{\mathcal{O}}(\theta_{2},\dots,\theta_{n},\theta_{1})\label{eq:cyclicaxiom}\\
\mbox{Kinematical poles:} &  & -i\mathop{\textrm{Res}}_{\theta=\theta'}F_{n+2}^{\mathcal{O}}(\theta+i\pi,\theta',\theta_{1},\dots,\theta_{n})=\label{eq:kinematicalaxiom}\\
 &  & \qquad\left(1-\prod_{k=1}^{n}S(\theta'-\theta_{k})\right)F_{n}^{\mathcal{O}}(\theta_{1},\dots,\theta_{n})\nonumber 
\end{eqnarray}
where $s_{\mathcal{O}}$ denotes the Lorentz spin of the operator
$\mathcal{O}$. There is also a further equation related to bound
states which we do not need in the sequel.

\subsection{Form factors in finite volume\label{finvolFF}}

A formalism that gives the exact quantum form factors to all orders
in $L^{-1}$ was introduced in \cite{Pozsgay:2007kn,Pozsgay:2007gx}.
The finite volume multi-particle states can be denoted 
\[
\vert\{I_{1},\dots,I_{n}\}\rangle_{L}
\]
where the $I_{k}$ are momentum quantum numbers, ordered as $I_{1}\geq\dots\geq I_{n}$
by convention. The corresponding energy levels are determined by the
Bethe-Yang equations 
\[
\mathrm{e}^{imL\sinh\tilde{\theta}_{k}}\prod_{l\neq k}S(\tilde{\theta}_{k}-\tilde{\theta}_{l})=1
\]
Defining the two-particle phase shift $\delta(\theta)$ by the relation
\begin{equation}
S(\theta)=-\mathrm{e}^{i\delta(\theta)}\label{deltadef}
\end{equation}
The derivative of $\delta$ will be denoted by 
\begin{equation}
\varphi(\theta)=\frac{d\delta(\theta)}{d\theta}\label{eq:phidef}
\end{equation}
due to unitarity, $\delta$ is an odd and $\varphi$ is an even function.
We can write 
\begin{equation}
Q_{k}(\tilde{\theta}_{1},\dots,\tilde{\theta}_{n})=mL\sinh\tilde{\theta}_{k}+\sum_{l\neq k}\delta(\tilde{\theta}_{k}-\tilde{\theta}_{l})=2\pi I_{k}\quad,\quad k=1,\dots,n\label{eq:betheyang}
\end{equation}
where the quantum numbers $I_{k}$ take integer/half-integer values
for odd/even numbers of particles respectively. Eqns. \eqref{eq:betheyang}
must be solved with respect to the particle rapidities $\tilde{\theta}_{k}$,
where the energy (relative to the finite volume vacuum state) can
be computed as 
\begin{equation}
\sum_{k=1}^{n}m\cosh\tilde{\theta}_{k}
\end{equation}
up to corrections which decay exponentially with $L$. The density
of $n$-particle states in rapidity space can be calculated as 
\begin{equation}
\rho(\theta_{1},\dots,\theta_{n})=\det\mathcal{J}^{(n)}\qquad,\qquad\mathcal{J}_{kl}^{(n)}=\frac{\partial Q_{k}(\theta_{1},\dots,\theta_{n})}{\partial\theta_{l}}\quad,\quad k,l=1,\dots,n\label{eq:byjacobian}
\end{equation}
The finite volume behavior of local matrix elements can be given as
\cite{Pozsgay:2007kn} 
\begin{eqnarray}
\langle\{I_{1}',\dots,I_{m}'\}\vert\mathcal{O}(0,0)\vert\{I_{1},\dots,I_{n}\}\rangle_{L} & = & \frac{F_{m+n}^{\mathcal{O}}(\tilde{\theta}_{m}'+i\pi,\dots,\tilde{\theta}_{1}'+i\pi,\tilde{\theta}_{1},\dots,\tilde{\theta}_{n})}{\sqrt{\rho(\tilde{\theta}_{1},\dots,\tilde{\theta}_{n})\rho(\tilde{\theta}_{1}',\dots,\tilde{\theta}_{m}')}}\nonumber \\
 & + & O(\mathrm{e}^{-\mu L})\label{eq:genffrelation}
\end{eqnarray}
where $\tilde{\theta}_{k}$ ($\tilde{\theta}_{k}'$) are the solutions
of the Bethe-Yang equations (\ref{eq:betheyang}) corresponding to
the state with the specified quantum numbers $I_{1},\dots,I_{n}$
($I_{1}',\dots,I_{n}'$) at the given volume $L$. The above relation
is valid provided there are no disconnected terms i.e. the left and
the right states do not contain particles with the same rapidity,
i.e. the sets $\left\{ \tilde{\theta}_{1},\dots,\tilde{\theta}_{n}\right\} $
and $\left\{ \tilde{\theta}_{1}',\dots,\tilde{\theta}_{m}'\right\} $
are disjoint.

It is easy to see that in the presence of nontrivial scattering there
are only two cases when exact equality of (at least some of) the rapidities
can occur \cite{Pozsgay:2007gx}: 
\begin{enumerate}
\item The two states are identical, i.e. $n=m$ and 
\[
\{I_{1}',\dots,I_{m}'\}=\{I_{1},\dots,I_{n}\}
\]
in which case the corresponding diagonal matrix element can be written
as a sum over all bipartite divisions of the set of the $n$ particles
involved (including the trivial ones when $A$ is the empty set or
the complete set $\{1,\dots,n\}$) 
\begin{equation}
\langle\{I_{1}\dots I_{n}\}|\mathcal{O}|\{I_{1}\dots I_{n}\}\rangle_{L}=\frac{\sum_{A\subset\{1,2,\dots n\}}\mathcal{F}(A)_{L}\rho(\{1,\dots,n\}\setminus A)_{L}}{\rho(\{1,\dots,n\})_{L}}+O(\mathrm{e}^{-\mu L})\label{diagff}
\end{equation}
where 
\[
\rho(\{k_{1},\dots,k_{r}\})_{L}=\rho(\tilde{\theta}_{k_{1}},\dots,\tilde{\theta}_{k_{r}})
\]
is the $r$-particle Bethe-Yang Jacobi determinant (\ref{eq:byjacobian})
involving only the $r$-element subset $1\leq k_{1}<\dots<k_{r}\leq n$
of the $n$ particles, and 
\begin{eqnarray*}
\mathcal{F}(\{k_{1},\dots,k_{r}\})_{L} & = & F_{2r}^{s}(\tilde{\theta}_{k_{1}},\dots,\tilde{\theta}_{k_{r}})\\
F_{2l}^{s}(\theta_{1},\dots,\theta_{l}) & = & \lim_{\epsilon\rightarrow0}F_{2l}^{\mathcal{O}}(\theta_{l}+i\pi+\epsilon,\dots,\theta_{1}+i\pi+\epsilon,\theta_{1},\dots,\theta_{l})
\end{eqnarray*}
is the so-called symmetric evaluation of diagonal multi-particle matrix
elements. 
\item Both states are parity symmetric states in the spin zero sector, i.e.
\begin{eqnarray*}
\{I_{1},\dots,I_{n}\} & \equiv & \{-I_{n},\dots,-I_{1}\}\\
\{I_{1}',\dots,I'_{m}\} & \equiv & \{-I'_{m},\dots,-I'_{1}\}
\end{eqnarray*}
Furthermore, both states must contain one (or possibly more, in a
theory with more than one species) particle of zero quantum number.
Writing $m=2k+1$ and $n=2l+1$ and defining 
\begin{eqnarray}
 &  & \mathcal{F}_{k,l}(\theta_{1}',\dots,\theta_{k}'|\theta_{1},\dots,\theta_{l})=\nonumber \\
 &  & \lim_{\epsilon\rightarrow0}F_{2k+2l+2}(i\pi+\theta_{1}'+\epsilon,\dots,i\pi+\theta_{k}'+\epsilon,i\pi-\theta_{k}'+\epsilon,\dots,i\pi-\theta_{1}'+\epsilon,\nonumber \\
 &  & i\pi+\epsilon,0,\theta_{1},\dots,\theta_{l},-\theta_{l},\dots,-\theta_{1})\label{eq:oddoddlimitdef}
\end{eqnarray}
the formula for the finite-volume matrix element takes the form 
\begin{eqnarray}
 &  & \langle\{I_{1}',\dots,I_{k}',0,-I_{k}',\dots,-I_{1}'\}|\mathcal{O}|\{I_{1},\dots,I_{l},0,-I_{l},\dots,-I_{1}\}\rangle_{L}\label{eq:oddoddlyrule}\\
 & = & \left(\rho_{2k+1}(\tilde{\theta}_{1}',\dots,\tilde{\theta}_{k}',0,-\tilde{\theta}_{k}',\dots,-\tilde{\theta}_{1}')\rho_{2l+1}(\tilde{\theta}_{1},\dots,\tilde{\theta}_{l},0,-\tilde{\theta}_{l},\dots,-\tilde{\theta}_{1})\right)^{-1/2}\nonumber \\
 &  & \times\Big[\mathcal{F}_{k,l}(\tilde{\theta}_{1}',\dots,\tilde{\theta}_{k}'|\tilde{\theta}_{1},\dots,\tilde{\theta}_{l})\nonumber \\
 &  & +mL\, F_{2k+2l}(i\pi+\tilde{\theta}_{1}',\dots,i\pi+\tilde{\theta}_{k}',i\pi-\tilde{\theta}_{k}',\dots,i\pi-\tilde{\theta}_{1}',\tilde{\theta}_{1},\dots,\tilde{\theta}_{l},-\tilde{\theta}_{l},\dots,-\tilde{\theta}_{1})\Big]\nonumber \\
 &  & +O(\mathrm{e}^{-\mu L})\nonumber 
\end{eqnarray}
 
\end{enumerate}

\subsection{The form factor expansion using finite volume regularization\label{subsec:2ptorganization}}

Using the finite volume description introduced in subsection \ref{finvolFF}
we can write 
\begin{equation}
\langle\mathcal{O}_{1}(x,t)\mathcal{O}_{2}(0)\rangle_{L}^{R}=\frac{1}{Z}\sum_{N,M}C_{NM}\label{Imn_def}
\end{equation}
where 
\begin{eqnarray}
C_{NM} & = & \sum_{I_{1}\dots I_{N}}\sum_{J_{1}\dots J_{M}}\bra{\{I_{1}\dots I_{N}\}}\mathcal{O}_{1}(0)\ket{\{J_{1}\dots J_{M}\}}_{L}\times\nonumber \\
 &  & \bra{\{J_{1}\dots J_{M}\}}\mathcal{O}_{2}(0)\ket{\{I_{1}\dots I_{N}\}}_{L}e^{i(P_{1}-P_{2})x}e^{-E_{1}(R-t)}e^{-E_{2}t}\label{cnm}
\end{eqnarray}
and $E_{1,2}$ and $P_{1,2}$ are the total energies and momenta of
the multi-particle states $\ket{\{I_{1}\dots I_{N}\}}_{L}$ and $\ket{\{J_{1}\dots J_{M}\}}_{L}$.
The task is to calculate the sum in finite volume and then take the
limit $L\rightarrow\infty$.

First we classify the contributions into different multi-particle
orders following the procedure in \cite{Essler:2009zz,Pozsgay:2010cr}.
Introducing two auxiliary variables $u$ and $v$ (at the end both
will be set to $1$): 
\begin{equation}
\langle\mathcal{O}_{1}(x,t)\mathcal{O}_{2}(0)\rangle_{L}^{R}=\frac{1}{Z}\sum_{N,M}u^{N}v^{M}C_{NM}\label{complete1}
\end{equation}
Similarly for the partition function 
\[
Z=\sum_{N}(uv)^{N}Z_{N}
\]
with $Z_{N}$ denoting the $N$-particle contribution. The inverse
of the partition function is expanded as 
\[
Z^{-1}=\sum_{N}(uv)^{N}\bar{Z}_{N}
\]
where 
\[
\bar{Z}_{0}=1\qquad\bar{Z}_{1}=-Z_{1}\qquad\bar{Z}_{2}=Z_{1}^{2}-Z_{2}
\]
Putting this together we can rewrite the expansion as 
\begin{equation}
\langle\mathcal{O}_{1}(x,t)\mathcal{O}_{2}(0)\rangle_{L}^{R}=\sum u^{N}v^{N}\tilde{D}_{NM}\label{complete}
\end{equation}
with 
\begin{equation}
\tilde{D}_{NM}=\sum_{l}C_{N-l,M-l}\bar{Z}_{l}\label{dnm}
\end{equation}
The first few nontrivial terms are given by 
\begin{equation}
\begin{split}\tilde{D}_{1M} & =C_{1M}-Z_{1}C_{0,M-1}\\
\tilde{D}_{2M} & =C_{2M}-Z_{1}C_{1,M-1}+(Z_{1}^{2}-Z_{2})C_{0,M-2}
\end{split}
\label{firstfewDtildes}
\end{equation}
In this way we produce a double series expansions in powers of the
variables $e^{-mt}$ and $e^{-m(R-t)}$. Since these variables are
independent, each quantity $\tilde{D}_{NM}$ must have a well-defined
$L\to\infty$ limit which we denote as 
\begin{equation}
D_{NM}=\lim_{L\to\infty}\tilde{D}_{NM}\label{dnm-uj}
\end{equation}
and we obtain that 
\begin{equation}
\langle\mathcal{O}_{1}(x,t)\mathcal{O}_{2}(0)\rangle^{R}=\lim_{L\to\infty}\langle\mathcal{O}_{1}(x,t)\mathcal{O}_{2}(0)\rangle_{L}^{R}=\sum_{N,M}D_{NM}\label{eq:2ptwithD}
\end{equation}
A similar reordering was also used for the expansion of the one-point
function in powers of $e^{-mR}$ \cite{Pozsgay:2007gx}, and for the
boundary one-point function in \cite{Takacs:2008ec}. It is evident
from \eqref{cnm} that the $D_{NM}$ with $N>M$ can be obtained from
those with $N<M$ after a trivial exchange of $t$ with $R-t$, $x$
with $-x$ and $\mathcal{O}_{1}$ with $\mathcal{O}_{2}$.

\section{The spectral expansion for finite temperature correlators\label{sec:The-spectral-expansion}}

To evaluate the finite temperature two-point function, it is necessary
to evaluate the summation over two sets of intermediate states. For
a given $C_{NM}$ this involves an $N$ and an $M$ particle state.
One can start with any of these; to simplify the calculations, it
is best to start with the one containing the smallest number or particles,
and do the other later. On the other hand, doing the calculation in
the reverse order allows one to cross-check the result \cite{Pozsgay:2010cr}.

To evaluate the first summation, a systematic method was given in
\cite{Pozsgay:2010cr} based on a multidimensional residue method.
Once this is done, all the singularities from the form factors are
tamed, and the second summation can be performed by a simple transition
from the discrete sum to an integral using the density of states.
Then, after assembling $\tilde{D}_{NM}$ using the lower $C_{N'M'}$
coefficients as in (\ref{dnm}), and taking the limit $L\rightarrow\infty$
the final formula for the contribution $D_{NM}$ can be obtained.
Another quick validity check of the calculation is provided by the
existence of the infinite volume limit.

\subsection{Converting sums to contour integrals\label{sub:Converting-sums-to}}

For sums over one-particles states $\ket{\{I\}}_{L}$ with quantum
number $I\in\mathbb{Z}$ we can substitute

\[
\sum_{I}\rightarrow\sum_{I}\oint_{C_{I}}\frac{d\theta}{2\pi}\frac{\rho_{1}(\theta)}{\mathrm{e}^{iQ_{1}(\theta)}-1}
\]
where 
\[
Q_{1}(\theta)=mL\sinh\theta\qquad\rho_{1}(\theta)=Q_{1}'(\theta)=mL\cosh\theta
\]
and $C_{I}$ are small closed curves surrounding the solution of 
\[
Q_{1}(\theta)=2\pi I
\]
in the complex $\theta$ plane.

\begin{figure}
\begin{centering}
\psfrag{Im}{$\Im m\theta$}\psfrag{Re}{$\Re e\theta$}\includegraphics[scale=0.9]{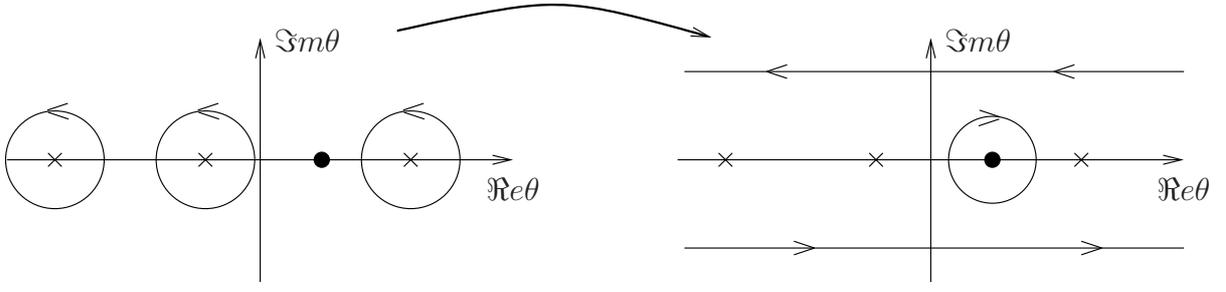}
\par\end{centering}

\caption{\label{fig:Contour-deformation-procedure} Contour deformation procedure.
The black dot shows a singularity not enclosed inside the contours
following from the spectral sum. }
\end{figure}

For two-particle sums over two-particle states $\ket{\{I_{1},I_{2}\}}_{L}$
with quantum numbers $I_{1},I_{2}\in\mathbb{Z}+\frac{1}{2}$ we can
use the multidimensional generalization of the residue theorem to
write 
\[
\sum_{I_{1}>I_{2}}\rightarrow\sum_{I_{1}>I_{2}}\oint\oint_{C_{I_{1}I_{2}}}\frac{d\theta_{1}}{2\pi}\frac{d\theta_{2}}{2\pi}\frac{\rho_{2}(\theta_{1},\theta_{2})}{\left(\mathrm{e}^{iQ_{1}(\theta_{1},\theta_{2})}+1\right)\left(\mathrm{e}^{iQ_{2}(\theta_{1},\theta_{2})}+1\right)}
\]
where $C_{I_{1}I_{2}}$ is a multi-contour (a direct product of two
curves in the variables $\theta_{1}$ and $\theta_{2}$) surrounding
the solution of 
\begin{eqnarray*}
Q_{1}(\theta_{1},\theta_{2}) & = & mL\sinh\theta_{1}+\delta(\theta_{1}-\theta_{2})=2\pi I_{1}\\
Q_{2}(\theta_{1},\theta_{2}) & = & mL\sinh\theta_{2}+\delta(\theta_{2}-\theta_{1})=2\pi I_{2}
\end{eqnarray*}
where due to the definition (\ref{deltadef}) $I_{1}$ and $I_{2}$
take half-integer values, and 
\[
\rho_{2}(\theta_{1},\theta_{2})=\det\left(\begin{array}{cc}
\frac{\partial Q_{1}}{\partial\theta_{1}} & \frac{\partial Q_{1}}{\partial\theta_{2}}\\
\frac{\partial Q_{2}}{\partial\theta_{1}} & \frac{\partial Q_{2}}{\partial\theta_{2}}
\end{array}\right)=m^{2}L^{2}\cosh\theta_{1}\cosh\theta_{2}+mL(\cosh\theta_{1}+\cosh\theta_{2})\varphi(\theta_{1}-\theta_{2})
\]
Since form factors vanish when any two of their arguments coincide,
we can extend the sum by adding the diagonal: 
\[
\sum_{I_{1}>I_{2}}\rightarrow\frac{1}{2}\left(\sum_{I_{1},I_{2}}-\sum_{I_{1}=I_{2}}\right)
\]
In the next step, the contours are joined together and opened into
straight lines, to a product contour whose components in each variable
enclose the real axis. However, this can only be done by including
other poles (apart from the ones needed for the state summations)
in the interior, which come from singularities of the $Q$-dependent
denominators and of the form factors. These must be classified and
subtracted. This procedure was discussed in some detail in \cite{Pozsgay:2010cr},
and for one complex variable it is illustrated in fig. \ref{fig:Contour-deformation-procedure}
(for more complex variable it must be performed in each variables
separately). We shall only outline it for the case of the $D_{22}$
contribution, because of the corrections we make to the previous calculation
performed in that paper.

\subsection{The $D_{22}$ contribution revisited\label{sub:The--contribution}}

The $D_{22}$ contribution is given by

\[
D_{22}=\lim_{L\to\infty}\left[C_{22}-Z_{1}C_{11}+\left(Z_{1}^{2}-Z_{2}\right)C_{00}\right]=\lim_{L\to\infty}\left[C_{22}-Z_{1}\tilde{D}_{11}-Z_{2}C_{00}\right]
\]
where

\begin{eqnarray*}
C_{22} & = & \sum_{I_{1}>I_{2}}\sum_{J_{1}>J_{2}}\langle\{I_{1},I_{2}\}|\mathcal{O}_{1}(0)|\{J_{1},J_{2}\}\rangle_{L}\langle\{J_{1},J_{2}\}|\mathcal{O}_{2}(0)|\{I_{1},I_{2}\}\rangle_{L}K_{t,x}^{\left(R\right)}(\tilde{\vartheta}_{1},\tilde{\vartheta}_{2}|\tilde{\vartheta}_{1}',\tilde{\vartheta}_{2}')
\end{eqnarray*}
with the notation

\begin{equation}
K_{t,x}^{\left(R\right)}(\vartheta_{1},\vartheta_{2}|\vartheta_{1}',\vartheta_{2}')=e^{imx\left(\sinh\vartheta_{1}+\sinh\vartheta_{2}-\sinh\vartheta_{1}'-\sinh\vartheta_{2}'\right)}e^{-m\left(R-t\right)\left(\cosh\vartheta_{1}+\cosh\vartheta_{2}\right)}e^{-mt\left(\cosh\vartheta_{1}'+\cosh\vartheta_{2}'\right)}\label{eq:4ptKRtx}
\end{equation}
and where \cite{Pozsgay:2010cr} 
\begin{align*}
Z_{1} & =mL\int\frac{\mathrm{d}\vartheta_{1}}{2\pi}\cosh\vartheta_{1}e^{-mR\cosh\vartheta_{1}}\\
Z_{2} & =\frac{1}{2}\iint\frac{\mathrm{d}\vartheta_{1}}{2\pi}\frac{\mathrm{d}\vartheta_{2}}{2\pi}\rho_{2}(\vartheta_{1},\vartheta_{2})e^{-mR\left(\cosh\vartheta_{1}+\cosh\vartheta_{2}\right)}-\frac{1}{2}\int\frac{\mathrm{d}\vartheta_{1}}{2\pi}\rho_{1}(\vartheta_{1})e^{-2mR\cosh\vartheta_{1}}\\
C_{00} & =\langle\mathcal{O}_{1}\rangle\langle\mathcal{O}_{2}\rangle\\
D_{11} & =\iint\frac{\mathrm{d}\vartheta_{1}}{2\pi}\frac{\mathrm{d}\vartheta_{2}}{2\pi}F_{2}^{\mathcal{O}_{1}}\left(\vartheta_{1}+i\pi,\vartheta_{2}\right)F_{2}^{\mathcal{O}_{2}}\left(\vartheta_{2}+i\pi,\vartheta_{1}\right)e^{imx\left(\sinh\vartheta_{1}-\sinh\vartheta_{2}\right)}e^{-m(R-t)\cosh\vartheta_{1}}e^{-mt\cosh\vartheta_{2}}\\
 & +\left[\langle\mathcal{O}_{1}\rangle F_{2s}^{\mathcal{O}_{2}}+\langle\mathcal{O}_{2}\rangle F_{2s}^{\mathcal{O}_{1}}\right]\int\frac{\mathrm{d}\vartheta_{1}}{2\pi}e^{-mR\cosh\vartheta_{1}}
\end{align*}
The rapidities are quantized by

\begin{align}
Q_{1}(\vartheta_{1},\vartheta_{2}) & =mL\sinh\vartheta_{1}+\delta(\vartheta_{1}-\vartheta_{2})=2\pi I_{1}\nonumber \\
Q_{2}(\vartheta_{1},\vartheta_{2}) & =mL\sinh\vartheta_{2}+\delta(\vartheta_{2}-\vartheta_{1})=2\pi I_{2}\label{eq:2ptBY}
\end{align}
and

\begin{align*}
Q_{1}'(\vartheta_{1}',\vartheta_{2}') & =mL\sinh\vartheta_{1}'+\delta(\vartheta_{1}'-\vartheta_{2}')=2\pi J_{1}\\
Q_{2}'(\vartheta_{1}',\vartheta_{2}') & =mL\sinh\vartheta_{2}'+\delta(\vartheta_{2}'-\vartheta_{1}')=2\pi J_{2}
\end{align*}
We perform the $J_{1},J_{2}$-sum first and separate it into a diagonal
and an off-diagonal piece:

\[
\sum_{J_{1}>J_{2}}=\left(\left\{ J_{1},J_{2}\right\} =\left\{ I_{1},I_{2}\right\} \mbox{ term}\right)+\sum_{J_{1}>J_{2}}\phantom{}'\,
\]
because the finite volume form factor expressions are different for
the two types of contributions. In the second term, the prime indicates
that the diagonal contributions are excluded.

\subsubsection{The diagonal piece}

This calculation is exactly the same as in \cite{Pozsgay:2010cr},
so we only highlight the main steps. Starting from 
\begin{eqnarray*}
C_{22}^{diag} & = & \sum_{I_{1}>I_{2}}\langle I_{1},I_{2}|\mathcal{O}_{1}(0)|I_{1},I_{2}\rangle\langle I_{1},I_{2}|\mathcal{O}_{2}(0)|I_{1},I_{2}\rangle K_{t,x}^{\left(R\right)}(\vartheta_{1},\vartheta_{2}|\vartheta_{1},\vartheta_{2})
\end{eqnarray*}
where

\begin{eqnarray*}
\langle I_{1},I_{2}|\mathcal{O}(0)|I_{1},I_{2}\rangle & = & \frac{F_{4s}^{\mathcal{O}}\left(\vartheta_{1},\vartheta_{2}\right)+\left(\rho_{1}(\vartheta_{1})+\rho_{1}(\vartheta_{2})\right)F_{2s}^{\mathcal{O}}+\rho_{2}(\vartheta_{1},\vartheta_{2})\langle\mathcal{O}\rangle}{\rho_{2}(\vartheta_{1},\vartheta_{2})}
\end{eqnarray*}
with

\begin{eqnarray*}
F_{2s}^{\mathcal{O}}\left(\vartheta\right) & = & F_{2}^{\mathcal{O}}\left(i\pi,0\right)\\
F_{4s}^{\mathcal{O}}\left(\vartheta_{1},\vartheta_{2}\right) & = & \lim_{\varepsilon\to0}F_{4}^{\mathcal{O}}\left(\vartheta_{1}+i\pi+\varepsilon,\vartheta_{2}+i\pi+\varepsilon,\vartheta_{2},\vartheta_{1}\right)
\end{eqnarray*}
Writing the sum in terms of contour integrals, after opening the contours
and performing the large $L$ limit the diagonal contribution becomes
\begin{eqnarray}
C_{22}^{diag} & = & \frac{1}{2}\iint\frac{\mathrm{d}\vartheta_{1}}{2\pi}\frac{\mathrm{d}\vartheta_{2}}{2\pi}\Big[F_{4s}^{\mathcal{O}_{1}}\left(\vartheta_{1},\vartheta_{2}\right)\langle\mathcal{O}_{2}\rangle\nonumber \\
 &  & +\, F_{4s}^{\mathcal{O}_{2}}\left(\vartheta_{1},\vartheta_{2}\right)\langle\mathcal{O}_{1}\rangle\Big]\, e^{-mR\left(\cosh\vartheta_{1}+\cosh\vartheta_{2}\right)}\nonumber \\
 & + & \frac{1}{2}\iint\frac{\mathrm{d}\vartheta_{1}}{2\pi}\frac{\mathrm{d}\vartheta_{2}}{2\pi}\frac{\left[\cosh\vartheta_{1}+\cosh\vartheta_{2}\right]^{2}}{\cosh\vartheta_{1}\cosh\vartheta_{2}}F_{2}^{\mathcal{O}_{1}}\left(i\pi,0\right)F_{2}^{\mathcal{O}_{2}}\left(i\pi,0\right)e^{-mR\left(\cosh\vartheta_{1}+\cosh\vartheta_{2}\right)}\nonumber \\
 & - & \int\frac{\mathrm{d}\vartheta_{1}}{2\pi}\left[F_{2}^{\mathcal{O}_{1}}\left(i\pi,0\right)\langle\mathcal{O}_{2}\rangle+F_{2}^{\mathcal{O}_{2}}\left(i\pi,0\right)\langle\mathcal{O}_{1}\rangle\right]e^{-m2R\cosh\vartheta_{1}}\nonumber \\
 & + & \iint\frac{\mathrm{d}\vartheta_{1}}{2\pi}\frac{\mathrm{d}\vartheta_{2}}{2\pi}mL\cosh\vartheta_{1}\left[F_{2}^{\mathcal{O}_{1}}\left(i\pi,0\right)\langle\mathcal{O}_{2}\rangle+F_{2}^{\mathcal{O}_{2}}\left(i\pi,0\right)\langle\mathcal{O}_{1}\rangle\right]e^{-mR\left(\cosh\vartheta_{1}+\cosh\vartheta_{2}\right)}\nonumber \\
 & + & Z_{2}C_{00}\label{eq:C22diag_largeL}
\end{eqnarray}

\subsubsection{The non-diagonal part}

In the non-diagonal part, one can use 
\[
\langle I_{1},I_{2}|\mathcal{O}(0)|J_{1},J_{2}\rangle=\frac{F_{4}^{\mathcal{O}}(\vartheta_{2}+i\pi,\vartheta_{1}+i\pi,\vartheta_{1}',\vartheta_{2}')}{\sqrt{\rho_{2}(\vartheta_{1},\vartheta_{2})\rho_{2}(\vartheta_{1}',\vartheta_{2}')}}
\]
to write 
\begin{eqnarray*}
C_{22}^{nondiag} & = & \sum_{I_{1}>I_{2}}\sum_{J_{1}>J_{2}}\phantom{}'\,\Bigg\{\frac{F_{4}^{\mathcal{O}_{1}}(\vartheta_{2}+i\pi,\vartheta_{1}+i\pi,\vartheta_{1}',\vartheta_{2}')F_{4}^{\mathcal{O}_{2}}(\vartheta_{1}+i\pi,\vartheta_{2}+i\pi,\vartheta_{2}',\vartheta_{1}')}{\rho_{2}(\vartheta_{1},\vartheta_{2})\rho_{2}(\vartheta_{1}',\vartheta_{2}')}\\
 &  & \times K_{t,x}^{\left(R\right)}(\vartheta_{1},\vartheta_{2}|\vartheta_{1}',\vartheta_{2}')\Bigg\}\\
 & = & \sum_{I_{1}>I_{2}}\frac{\tilde{C}_{22}}{\rho_{2}(\vartheta_{1},\vartheta_{2})}
\end{eqnarray*}
where 
\begin{eqnarray*}
\tilde{C}_{22} & = & \sum_{J_{1}>J_{2}}\phantom{}'\,\underset{\mathcal{C}_{J_{1}}\times\mathcal{C}_{J_{2}}}{\oint\oint}\frac{\mathrm{d}\vartheta_{1}'}{2\pi}\frac{\mathrm{d}\vartheta_{2}'}{2\pi}\Bigg\{\frac{F_{4}^{\mathcal{O}_{1}}(\vartheta_{2}+i\pi,\vartheta_{1}+i\pi,\vartheta_{1}',\vartheta_{2}')F_{4}^{\mathcal{O}_{2}}(\vartheta_{1}+i\pi,\vartheta_{2}+i\pi,\vartheta_{2}',\vartheta_{1}')}{\left[e^{iQ_{1}'(\vartheta_{1}',\vartheta_{2}')}+1\right]\left[e^{iQ_{2}'(\vartheta_{1}',\vartheta_{2}')}+1\right]}\\
 &  & \times K_{t,x}^{\left(R\right)}(\vartheta_{1},\vartheta_{2}|\vartheta_{1}',\vartheta_{2}')\Bigg\}
\end{eqnarray*}
and the prime denotes the omission of the $\{J_{1},J_{2}\}=\{I_{1},I_{2}\}$
term. We can substitute 
\[
\sum_{J_{1}>J_{2}}\phantom{}'\,\rightarrow\frac{1}{2}\sum_{J_{1},J_{2}}\phantom{}'\,
\]
since the form factors vanish when any two of their rapidity arguments
are identical.

Now we open the contours to encircle the real axis in $\vartheta_{1}'$
and $\vartheta_{2}'$. However, that brings more singularities inside
the contour whose contribution must then be subtracted. These can
be classified as follows: 
\begin{enumerate}
\item \textbf{Spurious QQ-poles.} There are two such terms, which come from
including the poles with $J_{1,2}=I_{1,2}$ or $J_{1,2}=I_{2,1}$.
Their contribution vanishes for $L\to\infty$ \cite{Pozsgay:2010cr},
hence the term 'spurious'. However, they must be included in the numerical
tests, therefore we provide their form in eqn. (\ref{eq:SQQ}). Note
that the form factors are not singular in this case, although their
limits in such points are direction dependent. 
\item \textbf{QF-poles.} In this case one of the integrations has a pole
from a form factor, and the other one from a $Q$-term: 
\[
\begin{array}{cccccccc}
\mathbf{QFI}: & \vartheta_{1}' & = & \vartheta_{1} & \& & Q_{2}'(\vartheta_{1},\vartheta_{2}') & = & 2\pi J_{2}\\
\mathbf{QFII}: & \vartheta_{1}' & = & \vartheta_{2} & \& & Q_{2}'(\vartheta_{2},\vartheta_{2}') & = & 2\pi J_{2}\\
\mathbf{QFIII:} & \vartheta_{2}' & = & \vartheta_{1} & \& & Q_{1}'(\vartheta_{1}',\vartheta_{1}) & = & 2\pi J_{1}\\
\mathbf{QFIV}: & \vartheta_{2}' & = & \vartheta_{2} & \& & Q_{1}'(\vartheta_{1}',\vartheta_{2}) & = & 2\pi J_{1}
\end{array}
\]

\item \textbf{FF poles.} In this case poles in both integrals come from
form factors: 
\[
\begin{array}{cccccc}
\mathbf{FFI:} & \vartheta_{1}' & = & \vartheta_{2}' & = & \vartheta_{1}\\
\mathbf{FFII:} & \vartheta_{1}' & = & \vartheta_{2}' & = & \vartheta_{2}
\end{array}
\]

\end{enumerate}
The poles of the form factors can be separated by introducing the
regular connected part $F_{4rc}$:

\begin{eqnarray}
F_{4}^{\mathcal{O}_{1}}(\vartheta_{2}+i\pi,\vartheta_{1}+i\pi,\vartheta_{1}',\vartheta_{2}') & = & \frac{A}{\vartheta_{2}-\vartheta_{1}'}+\frac{B}{\vartheta_{2}-\vartheta_{2}'}+\frac{C}{\vartheta_{1}-\vartheta_{1}'}+\frac{D}{\vartheta_{1}-\vartheta_{2}'}\nonumber \\
 &  & +F_{4rc}^{\mathcal{O}_{1}}(\vartheta_{2}+i\pi,\vartheta_{1}+i\pi|\vartheta_{1}',\vartheta_{2}')\nonumber \\
F_{4}^{\mathcal{O}_{2}}(\vartheta_{1}+i\pi,\vartheta_{2}+i\pi,\vartheta_{2}',\vartheta_{1}') & = & \frac{E}{\vartheta_{1}-\vartheta_{2}'}+\frac{F}{\vartheta_{1}-\vartheta_{1}'}+\frac{G}{\vartheta_{2}-\vartheta_{2}'}+\frac{H}{\vartheta_{2}-\vartheta_{1}'}\nonumber \\
 &  & +F_{4rc}^{\mathcal{O}_{2}}(\vartheta_{1}+i\pi,\vartheta_{2}+i\pi|\vartheta_{2}',\vartheta_{1}')\label{eq:F4rc_definition}
\end{eqnarray}
where 
\begin{eqnarray*}
A & = & i\left(S(\vartheta_{2}-\vartheta_{1})-S(\vartheta_{1}'-\vartheta_{2}')\right)F_{2}^{\mathcal{O}_{1}}(\vartheta_{1}+i\pi,\vartheta_{2}')\\
B & = & i\left(S(\vartheta_{1}'-\vartheta_{2}')S(\vartheta_{2}-\vartheta_{1})-1\right)F_{2}^{\mathcal{O}_{1}}(\vartheta_{1}+i\pi,\vartheta_{1}')\\
C & = & i\left(1-S(\vartheta_{2}-\vartheta_{1})S(\vartheta_{1}'-\vartheta_{2}')\right)F_{2}^{\mathcal{O}_{1}}(\vartheta_{2}+i\pi,\vartheta_{2}')\\
D & = & i\left(S(\vartheta_{1}'-\vartheta_{2}')-S(\vartheta_{2}-\vartheta_{1})\right)F_{2}^{\mathcal{O}_{1}}(\vartheta_{2}+i\pi,\vartheta_{1}')\\
E & = & i\left(S(\vartheta_{1}-\vartheta_{2})-S(\vartheta_{2}'-\vartheta_{1}')\right)F_{2}^{\mathcal{O}_{2}}(\vartheta_{2}+i\pi,\vartheta_{1}')\\
F & = & i\left(S(\vartheta_{2}'-\vartheta_{1}')S(\vartheta_{1}-\vartheta_{2})-1\right)F_{2}^{\mathcal{O}_{2}}(\vartheta_{2}+i\pi,\vartheta_{2}')\\
G & = & i\left(1-S(\vartheta_{1}-\vartheta_{2})S(\vartheta_{2}'-\vartheta_{1}')\right)F_{2}^{\mathcal{O}_{2}}(\vartheta_{1}+i\pi,\vartheta_{1}')\\
H & = & i\left(S(\vartheta_{2}'-\vartheta_{1}')-S(\vartheta_{1}-\vartheta_{2})\right)F_{2}^{\mathcal{O}_{2}}(\vartheta_{1}+i\pi,\vartheta_{2}')
\end{eqnarray*}
Using the above notation, the pole terms resulting from the form factors
can be obtained:

\begin{eqnarray}
 &  & F_{4}^{\mathcal{O}_{1}}(\vartheta_{2}+i\pi,\vartheta_{1}+i\pi,\vartheta_{1}',\vartheta_{2}')F_{4}^{\mathcal{O}_{2}}(\vartheta_{1}+i\pi,\vartheta_{2}+i\pi,\vartheta_{2}',\vartheta_{1}')=\nonumber \\
 &  & F_{4rc}^{\mathcal{O}_{1}}(\vartheta_{2}+i\pi,\vartheta_{1}+i\pi|\vartheta_{1}',\vartheta_{2}')F_{4rc}^{\mathcal{O}_{2}}(\vartheta_{1}+i\pi,\vartheta_{2}+i\pi|\vartheta_{2}',\vartheta_{1}')\nonumber \\
 &  & +F_{4rc}^{\mathcal{O}_{1}}(\vartheta_{2}+i\pi,\vartheta_{1}+i\pi|\vartheta_{1}',\vartheta_{2}')\left[\frac{E}{\vartheta_{1}-\vartheta_{2}'}+\frac{F}{\vartheta_{1}-\vartheta_{1}'}+\frac{G}{\vartheta_{2}-\vartheta_{2}'}+\frac{H}{\vartheta_{2}-\vartheta_{1}'}\right]\nonumber \\
 &  & +F_{4rc}^{\mathcal{O}_{2}}(\vartheta_{1}+i\pi,\vartheta_{2}+i\pi|\vartheta_{2}',\vartheta_{1}')\left[\frac{A}{\vartheta_{2}-\vartheta_{1}'}+\frac{B}{\vartheta_{2}-\vartheta_{2}'}+\frac{C}{\vartheta_{1}-\vartheta_{1}'}+\frac{D}{\vartheta_{1}-\vartheta_{2}'}\right]\nonumber \\
 &  & +\frac{AH}{\left(\vartheta_{2}-\vartheta_{1}'\right)^{2}}+\frac{BG}{\left(\vartheta_{2}-\vartheta_{2}'\right)^{2}}+\frac{CF}{\left(\vartheta_{1}-\vartheta_{1}'\right)^{2}}+\frac{DE}{\left(\vartheta_{1}-\vartheta_{2}'\right)^{2}}\nonumber \\
 &  & +\frac{AE+DH}{\left(\vartheta_{2}-\vartheta_{1}'\right)\left(\vartheta_{1}-\vartheta_{2}'\right)}+\frac{AF+CH}{\left(\vartheta_{2}-\vartheta_{1}'\right)\left(\vartheta_{1}-\vartheta_{1}'\right)}+\frac{AG+BH}{\left(\vartheta_{2}-\vartheta_{1}'\right)\left(\vartheta_{2}-\vartheta_{2}'\right)}\nonumber \\
 &  & +\frac{BE+DG}{\left(\vartheta_{2}-\vartheta_{2}'\right)\left(\vartheta_{1}-\vartheta_{2}'\right)}+\frac{BF+CG}{\left(\vartheta_{2}-\vartheta_{2}'\right)\left(\vartheta_{1}-\vartheta_{1}'\right)}+\frac{CE+DF}{\left(\vartheta_{1}-\vartheta_{1}'\right)\left(\vartheta_{1}-\vartheta_{2}'\right)}\label{eq:F4F4poleterms}
\end{eqnarray}
from which one can identify the terms giving $QF$ and $FF$ type
singularities. For the residue calculation, the formulas of Appendix
\ref{sec:Residue-evaluations} can be used. This results in certain
differences from the result derived in \cite{Pozsgay:2010cr}, where
too simplistic evaluation of residues resulted in some inaccuracies
in the end result.

Once the residues are calculated, in the case of the $QF$ terms a
further summation remains which must be converted into an integral.
It has the general form (here written for the case $QFI$):

\[
\sum_{J_{2}\neq I_{2}}\frac{G\left(\vartheta_{1},\vartheta_{2},\vartheta_{2}'\right)}{\left.\frac{\partial Q_{2}'(\vartheta_{1},\vartheta_{2}')}{\partial\vartheta_{2}'}\right|_{Q_{2}'=2\pi J_{2}}}
\]
where $\vartheta_{2}'$ is a solution to 
\[
Q_{2}'(\vartheta_{1},\vartheta_{2}')=2\pi J_{2}
\]
and the case $J_{2}=I_{2}$ was omitted since it is a spurious $QQ$
singularity. One can convert the $J_{2}$ summation into integrals
using the residue formula

\[
-\sum_{J_{2}\neq I_{2}}\underset{\mathcal{C}_{J_{2}}}{\oint}\frac{\mathrm{d}\vartheta_{2}'}{2\pi}\frac{G\left(\vartheta_{1},\vartheta_{2},\vartheta_{2}'\right)}{e^{iQ_{2}'(\vartheta_{1},\vartheta_{2}')}+1}
\]
Opening the contours and taking care to eliminate the contributions
resulting from possible poles of the function $G$ lying on the real
$\vartheta_{2}'$ axis:

\begin{eqnarray}
 &  & -\underset{\leftrightarrows}{\oint}\frac{\mathrm{d}\vartheta_{2}'}{2\pi}\frac{G\left(\vartheta_{1},\vartheta_{2},\vartheta_{2}'\right)}{e^{iQ_{2}'(\vartheta_{1},\vartheta_{2}')}+1}+\underset{\mathcal{C}_{\vartheta_{2}}}{\oint}\frac{\mathrm{d}\vartheta_{2}'}{2\pi}\frac{G\left(\vartheta_{1},\vartheta_{2},\vartheta_{2}'\right)}{e^{iQ_{2}'(\vartheta_{1},\vartheta_{2}')}+1}+\sum_{\mathrm{poles\, of\,}G}\underset{\mathcal{C}_{\vartheta_{2}^{*}}}{\oint}\frac{\mathrm{d}\vartheta_{2}'}{2\pi}\frac{G\left(\vartheta_{1},\vartheta_{2},\vartheta_{2}'\right)}{e^{iQ_{2}'(\vartheta_{1},\vartheta_{2}')}+1}\nonumber \\
 & = & -\underset{\leftrightarrows}{\oint}\frac{\mathrm{d}\vartheta_{2}'}{2\pi}\frac{G\left(\vartheta_{1},\vartheta_{2},\vartheta_{2}'\right)}{e^{iQ_{2}'(\vartheta_{1},\vartheta_{2}')}+1}-\frac{G\left(\vartheta_{1},\vartheta_{2},\vartheta_{2}\right)}{\left.\frac{\partial Q_{2}'(\vartheta_{1},\vartheta_{2}')}{\partial\vartheta_{2}'}\right|_{\vartheta_{2}'=\vartheta_{2}}}+\sum_{\mathrm{poles\, of\,}G}\underset{\mathcal{C}_{\vartheta_{2}^{*}}}{\oint}\frac{\mathrm{d}\vartheta_{2}'}{2\pi}\frac{G\left(\vartheta_{1},\vartheta_{2},\vartheta_{2}'\right)}{e^{iQ_{2}'(\vartheta_{1},\vartheta_{2}')}+1}\label{eq:sumtointegralforpoles}
\end{eqnarray}
where the second term corrects for the subtraction of the $J_{2}=I_{2}$
case and $\vartheta_{2}^{*}$ denotes the location of the poles of
$G$. The notation $\leftrightarrows$ corresponds to the straight
line contours enclosing the real axis as illustrated in fig. \ref{fig:Contour-deformation-procedure}.
The full results of the residue calculations are given in Appendix
\ref{sec:Pole-terms-for}.

The $J_{2}=I_{2}$ term typically is of order $O(1/L)$, except for
second order pole contributions. This results in the following contribution
to the $QF$ terms:
\begin{align}
 & F_{2}^{\mathcal{O}_{1}}(i\pi,0)F_{2}^{\mathcal{O}_{2}}(i\pi,0)K_{t,x}^{\left(R\right)}(\vartheta_{1},\vartheta_{2}|\vartheta_{2},\vartheta_{1})\nonumber \\
 & \times\left(\frac{\left[mL\cosh\vartheta_{1}-\varphi\left(\vartheta_{1}-\vartheta_{2}\right)\right]}{\left[mL\cosh\vartheta_{2}+\varphi\left(\vartheta_{2}-\vartheta_{1}\right)\right]}+\frac{\left[mL\cosh\vartheta_{2}-\varphi\left(\vartheta_{2}-\vartheta_{1}\right)\right]}{\left[mL\cosh\vartheta_{1}+\varphi\left(\vartheta_{1}-\vartheta_{2}\right)\right]}\right)\label{eq:criticaltermforclustering}
\end{align}
which is included in $QF6$ in (\ref{eq:QF6}). This term was omitted
by the calculation performed in \cite{Pozsgay:2010cr}; its presence
is critical for the cluster property.

\subsubsection{Performing the $I_{1},I_{2}$ sum and the large volume limit}

We can write 
\begin{eqnarray*}
C_{22}^{nondiag} & = & \sum_{I_{1}>I_{2}}\frac{\tilde{C}_{22}(\vartheta_{1},\vartheta_{2})}{\rho_{2}(\vartheta_{1},\vartheta_{2})}=\frac{1}{2}\left(\sum_{I_{1},I_{2}}-\sum_{I_{1}=I_{2}}\right)\frac{\tilde{C}_{22}(\vartheta_{1},\vartheta_{2})}{\rho_{2}(\vartheta_{1},\vartheta_{2})}=\\
 & = & \frac{1}{2}\sum_{I_{1},I_{2}}\underset{\mathcal{C}_{I_{1}}\times\mathcal{C}_{I_{2}}}{\oint\oint}\frac{\mathrm{d}\vartheta_{1}}{2\pi}\frac{\mathrm{d}\vartheta_{2}}{2\pi}\frac{\tilde{C}_{22}(\vartheta_{1},\vartheta_{2})}{\left[e^{iQ_{1}(\vartheta_{1},\vartheta_{2})}+1\right]\left[e^{iQ_{2}(\vartheta_{1},\vartheta_{2})}+1\right]}\\
 &  & +\frac{1}{2}\sum_{I_{1}=I_{2}}\underset{\mathcal{C}_{I_{1}}}{\oint}\frac{\mathrm{d}\vartheta_{1}}{2\pi}\frac{\tilde{C}_{22}(\vartheta_{1},\vartheta_{1})}{\left[e^{iQ_{1}(\vartheta_{1},\vartheta_{1})}+1\right]}\frac{\rho_{1}(\vartheta_{1})}{\rho_{2}(\vartheta_{1},\vartheta_{1})}
\end{eqnarray*}
Since $\tilde{C}_{22}$ doesn't have any pole we open the contours
in the usual way enclosing the real axis as illustrated in fig. \ref{fig:Contour-deformation-procedure}.
For the $L\to\infty$ it is necessary to examine the behavior of the
$Q$-functions:
\begin{eqnarray*}
iQ_{1}(\vartheta_{1}+i\varepsilon_{1},\vartheta_{2}+i\varepsilon_{2}) & = & imL\sinh(\vartheta_{1}+i\varepsilon_{1})+i\delta(\vartheta_{1}+i\varepsilon_{1}-\vartheta_{2}-i\varepsilon_{2})=\\
 & = & imL\sinh\vartheta_{1}\cos\varepsilon_{1}-mL\cosh\vartheta_{1}\sin\varepsilon_{1}+i\delta(\vartheta_{1}+i\varepsilon_{1}-\vartheta_{2}-i\varepsilon_{2})
\end{eqnarray*}
and similarly for $Q_{2}$ and $Q_{1,2}'$. This results in the following
limits:

\[
\lim_{L\to\infty}\frac{1}{e^{iQ_{i}\left(\vartheta_{1}+i\varepsilon_{1},\vartheta_{2}+i\varepsilon_{2}\right)}+1}=\begin{cases}
1, & \varepsilon_{i}\in\left[0,\pi\right]+2n\pi\\
0, & \varepsilon_{i}\in\left[\pi,2\pi\right]+2n\pi
\end{cases}
\]

\[
\lim_{L\to\infty}\frac{1}{e^{iQ_{1}\left(\vartheta_{1}+i\varepsilon_{1},\vartheta_{1}+i\varepsilon_{1}\right)}+1}=\begin{cases}
1, & \varepsilon_{1}\in\left[0,\pi\right]+2n\pi\\
0, & \varepsilon_{1}\in\left[\pi,2\pi\right]+2n\pi
\end{cases}
\]

\[
\lim_{L\to\infty}\frac{1}{e^{iQ_{i}'\left(\vartheta_{1}'+i\varepsilon_{1},\vartheta_{2}'+i\varepsilon_{2}\right)}+1}=\begin{cases}
1, & \varepsilon_{i}\in\left[0,\pi\right]+2n\pi\\
0, & \varepsilon_{i}\in\left[\pi,2\pi\right]+2n\pi
\end{cases}
\]
Therefore only the upper contours need to be kept, since all other
terms vanish exponentially for large $L$:

\[
\frac{1}{2}\iint\frac{\mathrm{d}\vartheta_{1}}{2\pi}\frac{\mathrm{d}\vartheta_{2}}{2\pi}\tilde{C}_{22}(\vartheta_{1}+i\varepsilon,\vartheta_{2}+i\varepsilon)-\frac{1}{2}\int\frac{\mathrm{d}\vartheta_{1}}{2\pi}\tilde{C}_{22}(\vartheta_{1}+i\varepsilon,\vartheta_{1}+i\varepsilon)\frac{\rho_{1}(\vartheta_{1}+i\varepsilon)}{\rho_{2}(\vartheta_{1}+i\varepsilon,\vartheta_{1}+i\varepsilon)}
\]
In addition, the integrals can be shifted to the real axis. However
this leads to singularities in the contribution like QF5 (\ref{eq:QF5})
due to the term containing 
\begin{equation}
\frac{K_{t,x}^{\left(R\right)}(\vartheta_{1},\vartheta_{2}|\vartheta_{1}',\vartheta_{1})\left[S(\vartheta_{1}-\vartheta_{2})-S(\vartheta_{1}'-\vartheta_{1})\right]}{\left[e^{iQ_{1}'(\vartheta_{1}',\vartheta_{1})}+1\right]\left(\vartheta_{1}-\vartheta_{1}'\right)}\label{eq:pvalueterms}
\end{equation}
which can be treated using the identity

\[
\frac{1}{x\pm i\varepsilon}=\mathcal{P}\frac{1}{x}\mp i\pi\delta\left(x\right)
\]

\subsection{End result for $D_{22}$\label{sub:End-result-for}}

The terms divergent as $L\rightarrow\infty$ drop out when including
the contribution $-Z_{1}\tilde{D}_{11}-Z_{2}C_{00}$. We can also
combine some terms by introducing the function 
\begin{eqnarray}
 &  & F_{4ss}^{\mathcal{O}}(\vartheta_{1}+i\pi,\vartheta_{2}+i\pi,\vartheta_{2}',\vartheta_{1}')=\label{eq:F4ssdef}\\
 & = & \frac{i\left(S(\vartheta_{1}-\vartheta_{2})-S(\vartheta_{2}'-\vartheta_{1}')\right)}{\vartheta_{1}-\vartheta_{2}'}F_{2}^{\mathcal{O}}(\vartheta_{2}+i\pi,\vartheta_{1}')\nonumber \\
 & + & \frac{i\left(1-S(\vartheta_{1}-\vartheta_{2})S(\vartheta_{2}'-\vartheta_{1}')\right)}{\vartheta_{2}-\vartheta_{2}'}F_{2}^{\mathcal{O}}(\vartheta_{1}+i\pi,\vartheta_{1}')+\frac{i\left(S(\vartheta_{2}'-\vartheta_{1}')-S(\vartheta_{1}-\vartheta_{2})\right)}{\vartheta_{2}-\vartheta_{1}'}F_{2}^{\mathcal{O}}(\vartheta_{1}+i\pi,\vartheta_{2}')\nonumber \\
 & + & F_{4rc}^{\mathcal{O}}(\vartheta_{1}+i\pi,\vartheta_{2}+i\pi|\vartheta_{2}',\vartheta_{1}')\nonumber 
\end{eqnarray}
The end result is 
\begin{eqnarray}
D_{22} & = & \frac{1}{4}\iiiint\frac{\mathrm{d}\vartheta_{1}}{2\pi}\frac{\mathrm{d}\vartheta_{2}}{2\pi}\frac{\mathrm{d}\vartheta_{1}'}{2\pi}\frac{\mathrm{d}\vartheta_{2}'}{2\pi}F_{4}^{\mathcal{O}_{1}}\left(\vartheta_{2}+i\pi,\vartheta_{1}+i\pi,\vartheta_{1}'+i\varepsilon,\vartheta_{2}'+i\varepsilon\right)\label{eq:d22final}\\
 &  & \times F_{4}^{\mathcal{O}_{2}}\left(\vartheta_{1}+i\pi,\vartheta_{2}+i\pi,\vartheta_{2}'+i\varepsilon,\vartheta_{1}'+i\varepsilon\right)K_{t,x}^{\left(R\right)}\left(\vartheta_{1},\vartheta_{2}|\vartheta_{1}'+i\varepsilon,\vartheta_{2}'+i\varepsilon\right)\nonumber \\
 & + & \iint\frac{\mathrm{d}\vartheta_{1}}{2\pi}\frac{\mathrm{d}\vartheta_{2}}{2\pi}\wp\int\frac{\mathrm{d}\vartheta_{1}'}{2\pi}\left\{ F_{4ss}^{\mathcal{O}_{1}}\left(\vartheta_{1}+i\pi,\vartheta_{2}+i\pi|\vartheta_{1}',\vartheta_{1}\right)F_{2}^{\mathcal{O}_{2}}\left(\vartheta_{2}+i\pi,\vartheta_{1}'\right)\right.\nonumber \\
 &  & \left.+F_{4ss}^{\mathcal{O}_{2}}\left(\vartheta_{1}+i\pi,\vartheta_{2}+i\pi|\vartheta_{1}',\vartheta_{1}\right)F_{2}^{\mathcal{O}_{1}}\left(\vartheta_{2}+i\pi,\vartheta_{1}'\right)\right\} K_{t,x}^{\left(R\right)}\left(\vartheta_{1},\vartheta_{2}|\vartheta_{1},\vartheta_{1}'\right)\nonumber \\
 & + & \iint\frac{\mathrm{d}\vartheta_{1}}{2\pi}\frac{\mathrm{d}\vartheta_{2}}{2\pi}\int\frac{\mathrm{d}\vartheta_{1}'}{2\pi}F_{2}^{\mathcal{O}_{1}}\left(\vartheta_{2}+i\pi,\vartheta_{1}'\right)F_{2}^{\mathcal{O}_{2}}\left(\vartheta_{2}+i\pi,\vartheta_{1}'\right)K_{t,x}^{\left(R\right)}\left(\vartheta_{1},\vartheta_{2}|\vartheta_{1},\vartheta_{1}'\right)\nonumber \\
 &  & \times\left[\left(1-S\left(\vartheta_{1}'-\vartheta_{1}\right)S\left(\vartheta_{1}-\vartheta_{2}\right)\right)\left(mx\cosh\vartheta_{1}-imt\sinh\vartheta_{1}\right)\right.\nonumber \\
 &  & \left.\underline{-\varphi\left(\vartheta_{1}'-\vartheta_{1}\right)S\left(\vartheta_{1}'-\vartheta_{1}\right)S\left(\vartheta_{1}-\vartheta_{2}\right)}\right]\nonumber \\
 & - & \underline{\iint\frac{\mathrm{d}\vartheta_{1}}{2\pi}\frac{\mathrm{d}\vartheta_{2}}{2\pi}F_{2}^{\mathcal{O}_{1}}\left(\vartheta_{2}+i\pi,\vartheta_{1}\right)F_{2}^{\mathcal{O}_{2}}\left(\vartheta_{2}+i\pi,\vartheta_{1}\right)K_{t,x}^{\left(R\right)}\left(\vartheta_{1},\vartheta_{2}|\vartheta_{1},\vartheta_{1}\right)}\nonumber \\
 & - & \iint\frac{\mathrm{d}\vartheta_{1}}{2\pi}\frac{\mathrm{d}\vartheta_{1}'}{2\pi}F_{2}^{\mathcal{O}_{1}}\left(\vartheta_{1}+i\pi,\vartheta_{1}'\right)F_{2}^{\mathcal{O}_{2}}\left(\vartheta_{1}+i\pi,\vartheta_{1}'\right)K_{t,x}^{\left(R\right)}\left(\vartheta_{1},\vartheta_{1}|\vartheta_{1},\vartheta_{1}'\right)\nonumber \\
 & + & \frac{1}{2}\iint\frac{\mathrm{d}\vartheta_{1}}{2\pi}\frac{\mathrm{d}\vartheta_{2}}{2\pi}\left[F_{4s}^{\mathcal{O}_{1}}\left(\vartheta_{1},\vartheta_{2}\right)\langle\mathcal{O}_{2}\rangle+F_{4s}^{\mathcal{O}_{2}}\left(\vartheta_{1},\vartheta_{2}\right)\langle\mathcal{O}_{1}\rangle\right]K_{t,x}^{\left(R\right)}\left(\vartheta_{1},\vartheta_{2}|\vartheta_{1},\vartheta_{2}\right)\nonumber \\
 & + & \underline{\iint\frac{\mathrm{d}\vartheta_{1}}{2\pi}\frac{\mathrm{d}\vartheta_{2}}{2\pi}F_{2}^{\mathcal{O}_{1}}\left(i\pi,0\right)F_{2}^{\mathcal{O}_{2}}\left(i\pi,0\right)K_{t,x}^{\left(R\right)}\left(\vartheta_{1},\vartheta_{2}|\vartheta_{1},\vartheta_{2}\right)}\nonumber \\
 & - & \int\frac{\mathrm{d}\vartheta_{1}}{2\pi}\left[F_{2}^{\mathcal{O}_{1}}\left(i\pi,0\right)\langle\mathcal{O}_{2}\rangle+F_{2}^{\mathcal{O}_{2}}\left(i\pi,0\right)\langle\mathcal{O}_{1}\rangle\right]K_{t,x}^{\left(R\right)}\left(\vartheta_{1},\vartheta_{1}|\vartheta_{1},\vartheta_{1}\right)\nonumber 
\end{eqnarray}
where $\mathcal{P}$ denotes a principal value integral, $K_{t,x}^{\left(R\right)}$
is defined in (\ref{eq:4ptKRtx}) and $F_{4s}$ is the so-called symmetric
evaluation of the form factor used in \cite{Pozsgay:2007gx}:

\begin{eqnarray*}
F_{4s}^{\mathcal{O}}\left(\vartheta_{1},\vartheta_{2}\right) & = & \lim_{\epsilon\rightarrow0}F^{\mathcal{O}}(\vartheta_{1}+i\pi+\epsilon,\vartheta_{2}+i\pi+\epsilon,\vartheta_{2},\vartheta_{1})
\end{eqnarray*}
Note that by introducing $F_{4ss}$ we combined the terms (\ref{eq:pvalueterms})
into the second integral, hence the need for the principal value.

In (\ref{eq:d22final}), the underlined pieces are the contributions
that are different from the earlier calculation performed in \cite{Pozsgay:2010cr}.
The first underlined term only corrects a typo in \cite{Pozsgay:2010cr},
where this piece was printed with the wrong sign. The second one comes
from the subtraction of poles of the integrand in (\ref{eq:sumtointegralforpoles})
and the careful evaluation of the principal value term (\ref{eq:pvalueterms}),
both of which occur in the manipulation of the QF5 contributions (\ref{eq:QF5}). 

The third underlined term plays a crucial role in the cluster property.
It is the left-over from the term 
\[
\frac{1}{2}\iint\frac{\mathrm{d}\vartheta_{1}}{2\pi}\frac{\mathrm{d}\vartheta_{2}}{2\pi}\frac{\left[\cosh\vartheta_{1}+\cosh\vartheta_{2}\right]^{2}}{\cosh\vartheta_{1}\cosh\vartheta_{2}}F_{2}^{\mathcal{O}_{1}}\left(i\pi,0\right)F_{2}^{\mathcal{O}_{2}}\left(i\pi,0\right)e^{-mR\left(\cosh\vartheta_{1}+\cosh\vartheta_{2}\right)}
\]
present in the diagonal contribution (\ref{eq:C22diag_largeL}), the
dependence on 
\[
\frac{\left[\cosh\vartheta_{1}+\cosh\vartheta_{2}\right]^{2}}{\cosh\vartheta_{1}\cosh\vartheta_{2}}=\frac{\cosh\vartheta_{1}}{\cosh\vartheta_{2}}+\frac{\cosh\vartheta_{2}}{\cosh\vartheta_{1}}+2
\]
is simplified by the inclusion of the contribution (\ref{eq:criticaltermforclustering}),
coming from the second order pole terms collected in QF6 (\ref{eq:QF6}).
In the large $L$ limit, the terms depending on the $\cosh$ ratios
cancel, leaving us with the last underlined piece in (\ref{eq:d22final}).
As mentioned before, one of the mistakes made in the evaluation of
$D_{22}$ in \cite{Pozsgay:2010cr} was the omission of this piece.

\subsection{The full two-point function up to $D_{22}$\label{sub:The-full-two-point}}

For completeness, we also give here the lower contributions to the
two-point function. These are exactly the same as in \cite{Pozsgay:2009pv,Pozsgay:2010cr},
so we do not give the derivations here. The calculations are almost
trivial with the exception of $D_{12}$, where one can use either
the derivations presented in \cite{Pozsgay:2010cr}, or follow the
steps outlined above, with slight modifications. The terms $D_{NM}$
with $N\leq M\leq2$ are

\begin{eqnarray}
D_{00} & = & \left\langle \mathcal{O}_{1}\right\rangle \left\langle \mathcal{O}_{2}\right\rangle \nonumber \\
D_{01} & = & \int\frac{d\vartheta_{1}}{2\pi}F_{1}^{\mathcal{O}_{1}}F_{1}^{\mathcal{O}_{2}}e^{-imx\sinh\vartheta_{1}-mt\cosh\vartheta_{1}}\nonumber \\
D_{02} & = & \frac{1}{2}\int\frac{d\vartheta_{1}}{2\pi}\frac{d\vartheta_{2}}{2\pi}F_{2}^{\mathcal{O}_{1}}\left(\vartheta_{1},\vartheta_{2}\right)F^{\mathcal{O}_{2}}\left(\vartheta_{2},\vartheta_{1}\right)e^{-imx\left(\sinh\vartheta_{1}+\sinh\vartheta_{2}\right)-mt\left(\cosh\vartheta_{1}+\cosh\vartheta_{2}\right)}\nonumber \\
D_{11} & = & \iint\frac{\mathrm{d}\vartheta_{1}}{2\pi}\frac{\mathrm{d}\vartheta_{2}}{2\pi}F_{2}^{\mathcal{O}_{1}}\left(\vartheta_{1}+i\pi,\vartheta_{2}\right)F_{2}^{\mathcal{O}_{2}}\left(\vartheta_{2}+i\pi,\vartheta_{1}\right)\nonumber \\
 &  & \times\mathrm{e}^{imx\left(\sinh\vartheta_{1}-\sinh\vartheta_{2}\right)}\mathrm{e}^{-m\left(R-t\right)\cosh\vartheta_{1}}\mathrm{e}^{-mt\cosh\vartheta_{2}}\nonumber \\
 &  & +\left[\langle\mathcal{O}_{1}\rangle F_{2s}^{\mathcal{O}_{2}}+\langle\mathcal{O}_{2}\rangle F_{2s}^{\mathcal{O}_{1}}\right]\int\frac{\mathrm{d}\vartheta_{1}}{2\pi}e^{-mR\cosh\vartheta_{1}}\nonumber \\
D_{12} & = & \frac{1}{2}\iiint\frac{\mathrm{d}\vartheta_{1}}{2\pi}\frac{\mathrm{d}\vartheta_{1}'}{2\pi}\frac{\mathrm{d}\vartheta_{2}'}{2\pi}F_{3}^{\mathcal{O}_{1}}\left(\vartheta_{1}+i\left(\pi+\varepsilon\right),\vartheta_{1}',\vartheta_{2}'\right)F_{3}^{\mathcal{O}_{2}}\left(\vartheta_{1}+i\left(\pi+\varepsilon\right),\vartheta_{2}',\vartheta_{1}'\right)\nonumber \\
 &  & \times K_{t,x}^{(R)}\left(\vartheta_{1}+i\varepsilon|\vartheta_{1}',\vartheta_{2}'\right)\nonumber \\
 & + & \iint\frac{\mathrm{d}\vartheta_{1}'}{2\pi}\frac{\mathrm{d}\vartheta_{2}'}{2\pi}\left[F_{1}^{\mathcal{O}_{1}}F_{3rc}^{\mathcal{O}_{2}}\left(\vartheta_{1}'+i\pi|\vartheta_{1}',\vartheta_{2}'\right)+F_{1}^{\mathcal{O}_{2}}F_{3rc}^{\mathcal{O}_{1}}\left(\vartheta_{1}'+i\pi|\vartheta_{1}',\vartheta_{2}'\right)\right]K_{t,x}^{(R)}\left(\vartheta_{1}'|\vartheta_{1}',\vartheta_{2}'\right)\nonumber \\
 & + & \iint\frac{\mathrm{d}\vartheta_{1}'}{2\pi}\frac{\mathrm{d}\vartheta_{2}'}{2\pi}2i\frac{F_{1}^{\mathcal{O}_{1}}F_{1}^{\mathcal{O}_{2}}}{\left(\vartheta_{1}'-\vartheta_{2}'\right)}\left(S\left(\vartheta_{1}'-\vartheta_{2}'\right)-1\right)K_{t,x}^{(R)}\left(\vartheta_{1}'|\vartheta_{1}',\vartheta_{2}'\right)\nonumber \\
 & + & \iint\frac{\mathrm{d}\vartheta_{1}'}{2\pi}\frac{\mathrm{d}\vartheta_{2}'}{2\pi}F_{1}^{\mathcal{O}_{1}}F_{1}^{\mathcal{O}_{2}}K_{t,x}^{(R)}\left(\vartheta_{1}'|\vartheta_{1}',\vartheta_{2}'\right)\left(S\left(\vartheta_{1}'-\vartheta_{2}'\right)-1\right)\left(mx\cosh\vartheta_{1}'+im\left(R-t\right)\sinh\vartheta_{1}'\right)\nonumber \\
 & - & \int\frac{\mathrm{d}\vartheta_{1}'}{2\pi}F_{1}^{\mathcal{O}_{1}}F_{1}^{\mathcal{O}_{2}}K_{t,x}^{(R)}\left(\vartheta_{1}'|\vartheta_{1}',\vartheta_{1}'\right)\label{eq:d00d01d02d11d12}
\end{eqnarray}
and $D_{22}$ is given in (\ref{eq:d22final}). In $D_{12}$ we defined
the regular connected form factor function $F_{3rc}$ via the following
separation of the kinematical pole terms 
\[
F_{3}^{\mathcal{O}}(\vartheta_{1}+i\pi,\vartheta_{1}',\vartheta_{2}')=\frac{i\left(1-S(\vartheta_{1}'-\vartheta_{2}')\right)F_{1}^{\mathcal{O}}}{\vartheta_{1}-\vartheta_{1}'}+\frac{i\left(S(\vartheta_{1}'-\vartheta_{2}')-1\right)F_{1}^{\mathcal{O}}}{\vartheta_{1}-\vartheta_{2}'}+F_{3rc}^{\mathcal{O}}(\vartheta_{1}+i\pi|\vartheta_{1}',\vartheta_{2}')
\]
and used the abbreviation 
\[
K_{t,x}^{\left(R\right)}(\vartheta_{1}|\vartheta_{1}',\vartheta_{2}')=e^{imx\left(\sinh\vartheta_{1}-\sinh\vartheta_{1}'-\sinh\vartheta_{2}'\right)}e^{-m\left(R-t\right)\cosh\vartheta_{1}}e^{-mt\left(\cosh\vartheta_{1}'+\cosh\vartheta_{2}'\right)}
\]
The other contributions $D_{NM}$ for $M>N$ can be obtained from
$D_{MN}$ by exchanging $\mathcal{O}_{1}$ with $\mathcal{O}_{2}$
and replacing $t\rightarrow R-t$, $x\rightarrow-x$.

\subsection{The symmetry of the $D_{22}$ term}

The relation between the coefficients $D_{NM}$ and $D_{MN}$ stated
above, when applied to $D_{22}$ leads to the property that $D_{22}$
must be symmetric under the following transformation: 
\begin{eqnarray*}
t & \to & R-t\\
\mathcal{O}^{1} & \leftrightarrow & \mathcal{O}^{2}\\
x & \to & -x
\end{eqnarray*}
This is the same as requiring that the result should be independent
of which two-particle summation is performed first. However, when
implementing such a transformation in (\ref{eq:d22final}), the signs
of the $\epsilon$ terms change, and therefore the contours must be
pulled back to their original positions. The contour deformation encounters
all the singularities on the real axis that were treated previously
in this section, so the appropriate residue contributions must be
computed. This computation is relegated to Appendix \ref{sec:Symmetry-of-D22},
where it is demonstrated that the required symmetry property indeed
holds, providing the first nontrivial test of the result (\ref{eq:d22final}).

\section{Numerical verification of the analytic results \label{sec:Numerical-verification-of}}

The goal of this section is to validate the $D_{22}$ formula numerically.
For this purpose we evaluated directly the sum for the two-particle
states and compare it with the result of the contour integrals. For
calculations we used the sinh-Gordon model with the Lagrangian density
\begin{equation}
\mathcal{L}=\frac{1}{2}\partial_{\mu}\Phi\partial^{\mu}\Phi-\frac{m^{2}}{g^{2}}\cosh g\Phi\label{eq:sinhGL}
\end{equation}
The model contains one massive particle, and its two-particle scattering
matrix is simple but nontrivial:
\[
S(\theta)=\frac{\sinh\theta-i\sin\frac{\pi B}{2}}{\sinh\theta+i\sin\frac{\pi B}{2}}
\]
where 
\[
B=\frac{2g^{2}}{8\pi+g^{2}}
\]
The nontrivial $S$-matrix is important, since our formula contains
the scattering matrix and its derivative in an essential way which
we would like to verify. For the fields in the correlator, we chose
the exponential operators 
\begin{equation}
e^{kg\Phi}\label{eq:expop}
\end{equation}
normalized to have vacuum expectation value $1$, since their form
factors are explicitly known \cite{Fring:1992pt,Koubek:1993ke}: 
\begin{align}
F_{n}^{(k)}(\theta_{1},\theta_{2},\dots,\theta_{n}) & =H_{n}\frac{P_{n}^{(k)}(x_{1},x_{2}\dots,x_{n})}{\prod\limits _{i<j}(x_{i}+x_{j})}\prod_{i<j}f(\theta_{i}-\theta_{j})\nonumber \\
 & x_{i}=e^{\theta_{i}}\label{eq:sinhgexpff}
\end{align}
where the polynomials $P_{n}^{(k)}$ are given by
\begin{eqnarray*}
P_{1}^{(k)} & = & [k]\\
P_{n}^{(k)} & = & [k]\det M^{(n)}(k)\qquad n>1\\
 &  & M_{ij}^{(n)}(k)=[i-j+k]\sigma_{2i-j}^{(n)}(x_{1},x_{2}\dots,x_{n})\quad i,j=1,\dots,n-1
\end{eqnarray*}
with
\[
H_{n}=\left(\frac{4\sin\pi B/2}{f(i\pi)}\right)^{n/2}\qquad\mbox{and}\qquad[n]=\frac{\sin\frac{n\pi B}{2}}{\sin\frac{\pi B}{2}}
\]
Furthermore,$\sigma_{l}^{(n)}$ denotes the elementary symmetric polynomials
of $n$ variables defined by
\begin{eqnarray*}
\prod_{i=1}^{n}(x+x_{i}) & = & \sum_{l=1}^{n}x^{n-l}\sigma_{l}^{(n)}(x_{1},\dots,x_{n})\\
\sigma_{l}^{(n)}\equiv0 &  & \mbox{if }\: l<0\mbox{ or }l>n
\end{eqnarray*}
and the minimal two-particle form factor is given by 
\begin{equation}
f(\theta)=\mathcal{N}\exp\left[8\int_{0}^{\infty}\frac{dx}{x}\sin^{2}\left(\frac{x(i\pi-\theta)}{2\pi}\right)\frac{\sinh\frac{xB}{4}\sinh(1-\frac{B}{2})\frac{x}{2}\sinh\frac{x}{2}}{\sinh^{2}x}\right]\label{eq:fmin}
\end{equation}
where 
\begin{equation}
\mathcal{N}=\exp\left[-4\int_{0}^{\infty}\frac{dx}{x}\frac{\sinh\frac{xB}{4}\sinh(1-\frac{B}{2})\frac{x}{2}\sinh\frac{x}{2}}{\sinh^{2}x}\right]\label{eq:NN}
\end{equation}

\subsection{Evaluating the two-particle sum}

Numerical evaluation of the sum is only possible at finite volume.
The factors $K_{t,x}^{\left(R\right)}$ decrease exponentially at
large rapidities, so it is possible to choose a rapidity cutoff and
restrict the summation up to the corresponding Bethe-Yang quantum
number. However for large volume this quantum number cutoff is still
too big and it is practically impossible to evaluate the four particle
sum. The compromise is to evaluate only the inner two particle sum
with fixed outer rapidities at moderate volume. This is enough to
check the validity of all the nontrivial contour deformations and
residue manipulation in the $D_{22}$ calculation. We can write

\[
C_{22}^{nondiag}=\sum_{I_{1}>I_{2}}\frac{\tilde{C}_{22}\left(\vartheta_{1},\vartheta_{2}\right)}{\rho_{2}\left(\vartheta_{1},\vartheta_{2}\right)}
\]
with 
\begin{eqnarray*}
\tilde{C}_{22}\left(\vartheta_{1},\vartheta_{2}\right) & = & \sum_{J_{1}>J_{2}}\phantom{}'\,\frac{F_{4}^{\mathcal{O}_{1}}\left(\vartheta_{2}+i\pi,\vartheta_{1}+i\pi,\vartheta_{1}',\vartheta_{2}'\right)F_{4}^{\mathcal{O}_{2}}\left(\vartheta_{1}+i\pi,\vartheta_{2}+i\pi,\vartheta_{2}',\vartheta_{1}'\right)}{\rho_{2}\left(\vartheta_{1}',\vartheta_{2}'\right)}\\
 &  & \;\times K_{t,x}^{\left(R\right)}\left(\vartheta_{1},\vartheta_{2},\vartheta_{1}',\vartheta_{2}'\right)
\end{eqnarray*}
and evaluate $\tilde{C}_{22}$ for some given value of $\vartheta_{1,2}$,
corresponding to a solution of the Bethe-Yang equations (\ref{eq:2ptBY})
with some quantum numbers $\{I_{1},I_{2}\}$.

The parameters for the evaluation can be chosen to help with the convergence
of the summation, while ensuring that the structure of the expression
tested remains general. The exponential operators (\ref{eq:expop})
in the Sinh-Gordon model can be parametrized by the number $k$ that
we chose for our evaluations as $k_{1}=2$ for $\mathcal{O}_{1}$
and $k_{2}=4$ for $\mathcal{O}_{2}$. The essential structure of
the formula does not depend on this choice. The space-time parameters
and the temperature were chosen as $mx=0.0$, $mt=0.4$, and $mR=0.8$.
Setting $mx$ to zero does not hide any important structure of the
equation, but makes the expression real and that helps in comparing
the results with the contour integrals. The sum was evaluated with
several values for the volume, the sinh-Gordon coupling constant and
the outer rapidities: 
\begin{align*}
mL & =\left(10,\:15,\:20,\:25,\:30\right)\\
B & =\left(0.1,\:0.2,\:0.3,\:0.4\:,0.55,\:0.7,\:0.9\right)\\
\left\{ I_{1},I_{2}\right\}  & \in\left\{ \left\{ \frac{5}{2},\frac{1}{2}\right\} ,\left\{ \frac{11}{2},\frac{-5}{2}\right\} ,\left\{ \frac{-5}{2},\frac{-21}{2}\right\} ,\left\{ \frac{7}{2},\frac{-7}{2}\right\} ,\left\{ \frac{1}{2},\frac{-1}{2}\right\} \right\} 
\end{align*}
The rapidity cutoff for the quantum numbers included in the sum was
chosen as $\vartheta=3.0,\:4.0,\:5.0,\:6.0$, and the numerical results
showed that for the value $\vartheta=6.0$ the discrete sum was evaluated
within a relative error of less than $10^{-14}$.

\subsection{Evaluating the contour integrals}

To compare the results of the contour integrals with the direct sum,
the calculation must be performed at the same volumes. In this regime
the exponential and power corrections in volume $L$ are not negligible,
so they must be taken into account. Exponential corrections come from
the integration on the contours going under the real axis, while power
corrections come from the total derivative contribution in the second
order pole calculation and from the $\left\{ J_{1},J_{2}\right\} =\left\{ I_{1},I_{2}\right\} $
point in the pole and the double integral contributions. The explicit
formulas can be found in the Appendix \ref{sec:Pole-terms-for}.

The integration contours run below and above the real axis, and it
is important to find a choice that is optimal for numerical evaluation.
The form factors and hence the integrands have poles on the real axis,
so it would be better to integrate as far from the real axis as possible.
However for imaginary parts of rapidities larger than $\frac{\pi}{2}$
the factor $K_{t,x}^{\left(R\right)}\left(\vartheta_{1},\vartheta_{2},\vartheta_{1}',\vartheta_{2}'\right)$
becomes oscillating and exponentially growing in the rapidity parameters
instead of decaying. Another issue is that the form factors and scattering
matrices are also evaluated at rapidities that lie out of the physical
strip. In the sinh-Gordon model the scattering matrix and hence the
minimal form factor have poles out of the physical strip, with imaginary
positions that are proportional to the coupling parameter $B$ \cite{Fring:1992pt}.
Therefore the contour must be chosen to lie between these poles on
the one hand and the poles on the real axis on the other hand. At
the same time it must run as far away from all singularities as possible,
and also to be closer to the real axis than $\frac{\pi}{2}$. For
small $B$ this leaves little space for the contours so they run relatively
close to the poles, resulting in a larger error in the numerical integration.
The integration itself was performed using Mathematica%
\footnote{Wolfram Research Inc, \emph{Mathematica}, version 8.0, Champaign Illinois,
2010. %
} and the Cuba library for multidimensional numerical integrations
\cite{Hahn:2004fe}.

\subsection{Comparing the results}

\begin{table}[h]

\begin{centering}
\begin{tabular}{|c||c|c|c|c|c|}
\hline 
$B\backslash mL$  & $10$  & $15$  & $20$  & $25$  & $30$ \tabularnewline
\hline 
\hline 
$0.1$  & $8.56\times10^{-8}$  & $1.03\times10^{-7}$  & $4.93\times10^{-9}$  & $3.29\times10^{-7}$  & $1.31\times10^{-8}$ \tabularnewline
\hline 
$0.2$  & $7.38\times10^{-10}$  & $3.53\times10^{-10}$  & $6.29\times10^{-11}$  & $9.74\times10^{-10}$  & $1.96\times10^{-10}$ \tabularnewline
\hline 
$0.3$  & $1.74\times10^{-10}$  & $1.13\times10^{-10}$  & $1.25\times10^{-10}$  & $1.26\times10^{-10}$  & $1.26\times10^{-10}$ \tabularnewline
\hline 
$0.4$  & $1.42\times10^{-10}$  & $1.42\times10^{-10}$  & $1.41\times10^{-10}$  & $1.4\times10^{-10}$  & $1.4\times10^{-10}$ \tabularnewline
\hline 
$0.55$  & $1.41\times10^{-10}$  & $1.42\times10^{-10}$  & $1.42\times10^{-10}$  & $1.42\times10^{-10}$  & $1.42\times10^{-10}$ \tabularnewline
\hline 
$0.7$  & $1.33\times10^{-10}$  & $1.34\times10^{-10}$  & $1.33\times10^{-10}$  & $1.33\times10^{-10}$  & $1.33\times10^{-10}$ \tabularnewline
\hline 
$0.9$  & $1.3\times10^{-10}$  & $1.3\times10^{-10}$  & $1.3\times10^{-10}$  & $1.3\times10^{-10}$  & $1.3\times10^{-10}$ \tabularnewline
\hline 
\end{tabular}
\par\end{centering}

\centering{}\caption{Relative error of the difference between the direct sum and the contour
integral evaluation of $\tilde{C}_{22}\left(\vartheta_{1},\vartheta_{2}\right)$
with $\left\{ I_{1},I_{2}\right\} =\left\{ \frac{7}{2},\frac{-7}{2}\right\} $
\label{tab:result}}
\end{table}

Table \ref{tab:result} shows the relative deviation between the direct
sum and the contour integral evaluation of $\tilde{C}_{22}\left(\vartheta_{1},\vartheta_{2}\right)$
with $\left\{ I_{1},I_{2}\right\} =\left\{ \frac{7}{2},\frac{-7}{2}\right\} $.
Note that the relative error decreases as $B$ grows which can be
understood from the conditions for the choice of the integration contour
mentioned above. Based on the above understanding of the deviations
in the relative errors for different parameters, and the fact that
this pattern of dependence was the same for every value of $I_{1},I_{2}$
we checked, it can be inferred that the difference of the sum and
the contour integration is only due to the numerical errors of integration.

To provide a further support for this conclusion, the above numerical
test was repeated for $C_{12}$. The formula of $C_{12}$ is derived
in two independent ways in \cite{Pozsgay:2010cr} (depending on whether
the one-particle or the two-particle summation is performed first),
and therefore its validity is quite certain even without a numerical
test. As in the case of $C_{22}$, let us denote by $\tilde{C}_{12}\left(\vartheta_{1}\right)$
the result of performing the two-particle summation first with fixed
rapidity of the one-particle state. Table \ref{tab:result2} shows
the relative deviation between the direct sum and the contour integral
evaluation of $\tilde{C}_{12}\left(\vartheta_{1}\right)$ with $I_{1}=17$
as the Bethe-Yang quantum number of the one-particle state. The relative
deviation has the same pattern as for $\tilde{C}_{22}\left(\vartheta_{1},\vartheta_{2}\right)$,
and is essentially of the same magnitude. Therefore this evaluation
gives an independent support for the assertion that the deviations
are caused by errors of numerical integration.

As the derivation of $D_{22}$ from $\tilde{C}_{22}\left(\vartheta_{1},\vartheta_{2}\right)$
is almost trivial, the above numerical tests also confirm the details
of our analytic result for $D_{22}$.

\begin{table}[h]

\begin{centering}
\begin{tabular}{|c||c|c|c|c|c|}
\hline 
$B\backslash mL$  & $10$  & $15$  & $20$  & $25$  & $30$ \tabularnewline
\hline 
\hline 
$0.1$  & $2.03\times10^{-6}$  & $3.56\times10^{-6}$  & $3.41\times10^{-7}$  & $1.56\times10^{-7}$  & $8.81\times10^{-8}$ \tabularnewline
\hline 
$0.2$  & $7.5\times10^{-9}$  & $5.03\times10^{-9}$  & $1.97\times10^{-10}$  & $1.52\times10^{-9}$  & $7.73\times10^{-10}$ \tabularnewline
\hline 
$0.3$  & $1.48\times10^{-10}$  & $7.76\times10^{-11}$  & $3.1\times10^{-11}$  & $6.22\times10^{-11}$  & $3.13\times10^{-11}$ \tabularnewline
\hline 
$0.4$  & $2.35\times10^{-11}$  & $5.29\times10^{-11}$  & $5.81\times10^{-11}$  & $6.77\times10^{-11}$  & $6.59\times10^{-11}$ \tabularnewline
\hline 
$0.55$  & $5.87\times10^{-11}$  & $6.61\times10^{-11}$  & $6.87\times10^{-11}$  & $6.9\times10^{-11}$  & $6.96\times10^{-11}$ \tabularnewline
\hline 
$0.7$  & $2.96\times10^{-11}$  & $5.32\times10^{-11}$  & $6.02\times10^{-11}$  & $6.25\times10^{-11}$  & $6.38\times10^{-11}$ \tabularnewline
\hline 
$0.9$  & $6.01\times10^{-11}$  & $6.48\times10^{-11}$  & $6.52\times10^{-11}$  & $6.51\times10^{-11}$  & $6.52\times10^{-11}$ \tabularnewline
\hline 
\end{tabular}
\par\end{centering}

\caption{Relative error of the difference between the direct sum and the contour
integral evaluation of $\tilde{C}_{12}\left(\vartheta_{1}\right)$
with $I_{1}=17$ \label{tab:result2}}
\end{table}

\section{The cluster property of the two-point function\label{sec:The-cluster-property}}

Another important test of the results is provided by checking that
the two-point function has the cluster property 
\[
\langle\mathcal{O}_{1}(x,t)\mathcal{O}_{2}(0,0)\rangle^{R}\sim\langle\mathcal{O}_{1}(0,0)\rangle^{R}\langle\mathcal{O}_{2}(0,0)\rangle^{R}
\]
when the spatial separation $x$ grows large. Using the expansion
up to $D_{22}$ one can write

\begin{eqnarray*}
\langle\mathcal{O}_{1}(x,t)\mathcal{O}_{2}(0,0)\rangle^{R} & = & \sum_{N,M}D_{NM}\\
 & = & D_{00}+D_{01}+D_{10}+D_{11}+D_{02}+D_{20}+D_{12}+D_{21}+D_{22}+\dots
\end{eqnarray*}
For $mx\gg1$ the terms containing 
\[
e^{imx\left(\sum_{k}\sinh\vartheta_{k}-\sum_{l}\sinh\vartheta_{l}'\right)}
\]
oscillate very fast, and therefore the support of the (multiple) rapidity
integrals is restricted to the zero measure set 
\[
\sum_{k}\sinh\vartheta_{k}=\sum_{l}\sinh\vartheta_{l}'
\]
and the integral vanishes. Although this argument looks simple, there
is a possible problem. Namely, the argument only works if the integrands
of the $x$-dependent terms are all regular. A nontrivial example
is the term 
\[
\frac{1}{2}\iint\frac{\mathrm{d}\vartheta_{1}'}{2\pi}\frac{\mathrm{d}\vartheta_{2}'}{2\pi}\left\{ 2i\frac{F_{1}^{\mathcal{O}_{1}}F_{1}^{\mathcal{O}_{2}}}{\left(\vartheta_{1}'-\vartheta_{2}'\right)}\left(S(\vartheta_{1}'-\vartheta_{2}')-1\right)K_{t,x}^{(R)}(\vartheta_{1}'|\vartheta_{1}',\vartheta_{2}')+\left(\vartheta_{1}'\leftrightarrow\vartheta_{2}'\right)\right\} 
\]
in the contribution $D_{12}$ (cf. eqn. (\ref{eq:d00d01d02d11d12})),
which is in fact regular at $\vartheta_{1}'=\vartheta_{2}'$ when
the two terms inside the braces are added together. In the case of
the principal value integral in $D_{22}$ in eqn. (\ref{eq:d22final}),
the regularity of the integrand is ensured by the principal value
prescription itself.

As a result, one only needs to examine the terms that are $x$-independent.
We denote these by putting a bar over the respective contribution
$D_{NM}$ and they read:

\begin{eqnarray*}
\bar{D}_{00} & = & \langle\mathcal{O}_{1}\rangle\langle\mathcal{O}_{2}\rangle\\
\bar{D}_{01} & = & \bar{D}_{10}=\bar{D}_{02}=\bar{D}_{20}=0\\
\bar{D}_{11} & = & \left[\langle\mathcal{O}_{1}\rangle F_{2}^{\mathcal{O}_{2}}\left(i\pi,0\right)+\langle\mathcal{O}_{2}\rangle F_{2}^{\mathcal{O}_{1}}\left(i\pi,0\right)\right]\int\frac{\mathrm{d}\vartheta_{1}}{2\pi}e^{-mR\cosh\vartheta_{1}}\\
\bar{D}_{12} & = & \bar{D}_{21}=0\\
\bar{D}_{22} & = & +\frac{1}{2}\iint\frac{\mathrm{d}\vartheta_{1}}{2\pi}\frac{\mathrm{d}\vartheta_{2}}{2\pi}\Big[F_{4s}^{\mathcal{O}_{1}}\left(\vartheta_{1},\vartheta_{2}\right)\langle\mathcal{O}_{2}\rangle+F_{4s}^{\mathcal{O}_{2}}\left(\vartheta_{1},\vartheta_{2}\right)\langle\mathcal{O}_{1}\rangle\\
 &  & \;\times e^{-mR\left(\cosh\vartheta_{1}+\cosh\vartheta_{2}\right)}\\
 &  & +\iint\frac{\mathrm{d}\vartheta_{1}}{2\pi}\frac{\mathrm{d}\vartheta_{2}}{2\pi}F_{2}^{\mathcal{O}_{1}}\left(i\pi,0\right)F_{2}^{\mathcal{O}_{2}}\left(i\pi,0\right)e^{-mR\left(\cosh\vartheta_{1}+\cosh\vartheta_{2}\right)}\\
 &  & -\int\frac{\mathrm{d}\vartheta_{1}}{2\pi}\left[F_{2}^{\mathcal{O}_{1}}\left(i\pi,0\right)\langle\mathcal{O}_{2}\rangle+F_{2}^{\mathcal{O}_{2}}\left(i\pi,0\right)\langle\mathcal{O}_{1}\rangle\right]e^{-2mR\cosh\vartheta_{1}}
\end{eqnarray*}
The one-point function up to two-particle order is \cite{Pozsgay:2007gx}:

\begin{eqnarray*}
\langle\mathcal{O}\rangle^{R} & = & \langle\mathcal{O}\rangle+\int\frac{\mathrm{d}\vartheta_{1}}{2\pi}F_{2}^{\mathcal{O}}\left(i\pi,0\right)e^{-mR\cosh\vartheta_{1}}-\int\frac{\mathrm{d}\vartheta_{1}}{2\pi}F_{2}^{\mathcal{O}}\left(i\pi,0\right)e^{-2mR\cosh\vartheta_{1}}\\
 &  & +\frac{1}{2}\iint\frac{\mathrm{d}\vartheta_{1}}{2\pi}\frac{\mathrm{d}\vartheta_{2}}{2\pi}F_{4s}^{\mathcal{O}}\left(\vartheta_{1},\vartheta_{2}\right)e^{-mR\left(\cosh\vartheta_{1}+\cosh\vartheta_{2}\right)}+O\left(\mathrm{e}^{-3mR}\right)
\end{eqnarray*}
As a result one obtains that 
\[
\bar{D}_{00}+\bar{D}_{11}+\bar{D}_{22}=\langle\mathcal{O}_{1}\rangle^{R}\langle\mathcal{O}_{2}\rangle^{R}+O\left(\mathrm{e}^{-3mR}\right)
\]
and therefore the cluster property is satisfied to the given order.
Note that the same argument shows that the formula for $D_{22}$ derived
in \cite{Pozsgay:2010cr} violates the cluster property, providing
another argument that it needs to be corrected.

\section{Conclusions\label{sec:Conclusions}}

In this work we revisited the finite volume regularization of thermal
correlators introduced in \cite{Pozsgay:2010cr}, which is based on
the finite volume form factor formalism \cite{Pozsgay:2007kn,Pozsgay:2007gx}.
We have shown that the original results for the two-particle--two-particle
contribution $D_{22}$ need to be slightly corrected, and presented
a modified prescription for evaluation of residue contributions. As
a result, we now have the expansion up to all terms involving intermediate
states with no more than two particles. In addition, the result for
the nontrivial terms $D_{12}$ and $D_{22}$ was cross-checked with
a numerical evaluation of the finite volume summation over the intermediate
multi-particle states. It was also established that the correlation
function given by the final formulas (\ref{eq:d22final}) and (\ref{eq:d00d01d02d11d12})
satisfies the cluster property. In addition, it was shown to possess
a symmetry property which follows from the general structure of the
spectral expansion. 

The formalism presented here can be extended to compute any higher
correction in the series. However, the calculation of $D_{22}$ is
already very tedious, and it is expected to become even longer for
higher terms. In view of potential applications to condensed matter
systems, it is likely that the present evaluation would suffice for
most of the cases. 

Another reason why the result evaluated up to $D_{22}$ is interesting
is that this is the part which generalizes to non-integrable field
theories. Breaking integrability in general allows inelastic processes;
however, below the inelastic threshold the finite volume levels can
still be described using only the elastic phase shift and the same
quantization conditions as in (\ref{eq:2ptBY}) \cite{Luscher:1986pf,Luscher:1990ux}.
Therefore the present computation can be extended self-consistently
whenever the states dominating the spectral expansion are below the
inelastic threshold.

On the other hand, it remains an open question whether the full expansion
for the finite temperature two-point function in integrable models
can be recast in a form similar to the expression conjectured by Leclair
and Mussardo \cite{Leclair:1999ys}. Such an expression would represent
a partial re-summation of the series, expressing the correlator in
terms of Fermi-Dirac distribution for the dressed particles corresponding
to the representation of the system as a free gas of quasi-particles
under the thermodynamic Bethe Ansatz \cite{Zamolodchikov:1989cf}.
It is known that this re-summation is possible for the one-point function
\cite{Pozsgay:2007gx,Pozsgay:2010xd}, even for the case of operators
located on a system boundary \cite{Takacs:2008ec}. In that respect,
the explicit dependence of $D_{22}$ on the two-particle $S$-matrix
(cf. the first underlined term in eqn. (\ref{eq:d22final})) does
not bode well. The TBA equation contains only the derivative of the
phase-shift $\varphi$, so any dressing term from there is only expected
to depend only on $\varphi$. Some partial integration tricks can
be performed to shift this dependence around, but we have found no
way of eliminating it. On the other hand, it was noticed in \cite{Pozsgay:2010cr}
that the $D_{1n}$ contributions allow some re-summation by dressing
the contributions $D_{0n-1}$. There is a possibility that a redefinition
of the form factor terms could help, i.e. if one used some definition
for the desingularized form factors different from the $F_{rc}$ or
$F_{ss}$. At present it is not known how to accomplish this; however,
there is still hope for recovering some expression similar to the
original Leclair-Mussardo conjecture. Evaluation of some higher order
corrections could shed light on the structure of the series, and the
experience gained in the present work opens the way to performing
these calculations, armed with a numerical method to verify the results
of the complicated analytic manipulations. We hope to return to this
line of thought in the near future.

The present calculation was performed for a theory with a single massive
particles. Adding more particles to the spectrum is rather straightforward
as long as the scattering remains diagonal; it is only necessary to
add particle species labels in appropriate places. For non-diagonal
scattering theories, recent progress has made almost all matrix elements
available in finite volume \cite{Feher:2011aa,Feher:2011fn,Palmai:2011kb},
except for matrix elements involving disconnected pieces when the
states involved in the matrix element are subject to non-diagonal
scattering. However, more recently we have solved the issue for two-particle
states%
\footnote{In fact, the paper \cite{Palmai:2012pq} presents a conjectured solution
for any number of particles, but only the two-particle case is backed
up by (very strong) numerical evidence.%
} \cite{Palmai:2012pq}, which means that the series presented here
can be evaluated for general integrable field theories, including
those with non-diagonal scattering such as the sine-Gordon or O(3)
$\sigma$ models \cite{zam-zam}.

\subsubsection*{Acknowledgments}

We are grateful to Balázs Pozsgay for useful discussions and valuable
comments on the manuscript. GT was partially supported by the Hungarian
OTKA grants K75172 and K81461.

\appendix
\makeatletter 
\renewcommand{\theequation}{\hbox{\normalsize\Alph{section}.\arabic{equation}}} \@addtoreset{equation}{section} \renewcommand{\thefigure}{\hbox{\normalsize\Alph{section}.\arabic{figure}}} \@addtoreset{figure}{section} \renewcommand{\thetable}{\hbox{\normalsize\Alph{section}.\arabic{table}}} \@addtoreset{table}{section} \makeatother

\section{Residue evaluations \label{sec:Residue-evaluations}}

\subsection{First order poles}

In this case, the two-dimensional residue formula

\[
\underset{\mathcal{C}_{a}\times\mathcal{C}_{b}}{\oint\oint}\frac{\mathrm{d}z_{1}}{2\pi i}\frac{\mathrm{d}z_{2}}{2\pi i}\frac{g(z_{1},z_{2})}{f_{1}(z_{1},z_{2})f_{2}(z_{1},z_{2})}=\frac{g(a,b)}{\left.\det\left(\frac{\partial f_{i}}{\partial z_{j}}\right)\right|_{(z_{1},z_{2})=(a,b)}}
\]
can be applied directly as

\[
\underset{\mathcal{C}_{a}\times\mathcal{C}_{J}}{\oint\oint}\frac{\mathrm{d}z_{1}}{2\pi i}\frac{\mathrm{d}z_{2}}{2\pi i}\frac{g(z_{1},z_{2})}{\left[a-z_{1}\right]\left[e^{iQ_{2}(z_{1},z_{2})}+1\right]}=\frac{g(a,z_{2}^{*})}{\left.i\frac{\partial Q_{2}(a,z_{2})}{\partial z_{2}}\right|_{z_{2}=z_{2}^{*}}}
\]
where the $(a-z_{1})$ is the pole term coming from the appropriate
form factor, and $z_{2}^{*}$ is the root of 
\[
e^{iQ_{2}(a,z_{2}^{*})}+1=0
\]

\subsection{Second order poles}

These can be evaluated by successive integration, performing first
the integral over the second order pole coming from the form factor.
The fundamental formula to use is

\begin{equation}
\underset{z=a}{\mathrm{Res}}\frac{h(z)}{g(z)}=\frac{2h'(a)}{g^{''}(a)}-\frac{2g^{'''}(a)}{3\left[g^{''}(a)\right]^{2}}h(a)\label{eq:2ndorderpolefundamental}
\end{equation}
where for a second order pole $g(a)=g'(a)=0$ but $g^{''}(a)\neq0$.
The terms we need to evaluate have the form

\[
\underset{\mathcal{C}_{a}\times\mathcal{C}_{J}}{\oint\oint}\frac{\mathrm{d}z_{1}}{2\pi i}\frac{\mathrm{d}z_{2}}{2\pi i}\frac{g(z_{1},z_{2})}{\left[a-z_{1}\right]^{2}\left[e^{iQ_{2}(z_{1},z_{2})}+1\right]}
\]
Performing the $z_{1}$ integral leads to

\begin{align*}
 & \underset{\mathcal{C}_{J}}{\oint}\frac{\mathrm{d}z_{2}}{2\pi i}\left\{ \frac{\left.\frac{\partial g(z_{1},z_{2})}{\partial z_{1}}\right|_{z_{1}=a}}{\left[e^{iQ_{2}(z_{1},z_{2})}+1\right]}-\frac{g(a,z_{2})e^{iQ_{2}(a,z_{2})}\left.i\frac{\partial Q(z_{1},z_{2})}{\partial z_{1}}\right|_{z_{1}=a}}{\left[e^{iQ_{2}(z_{1},z_{2})}+1\right]^{2}}\right\} \\
 & =\frac{i\left.\frac{\partial g(z_{1},z_{2})}{\partial z_{1}}\right|_{z_{1}=a}}{\left.\frac{\partial Q(z_{1},z_{2}^{*})}{\partial z_{1}}\right|_{z_{1}=a}}-\underset{\mathcal{C}_{J}}{\oint}\frac{\mathrm{d}z_{2}}{2\pi i}\frac{g(a,z_{2})e^{iQ_{2}(a,z_{2})}\left.i\frac{\partial Q(z_{1},z_{2})}{\partial z_{1}}\right|_{z_{1}=a}}{\left[e^{iQ_{2}(z_{1},z_{2})}+1\right]^{2}}
\end{align*}
For the second integral, introducing the notation $l(z_{1},z_{2})=g(z_{1},z_{2})i\frac{\partial Q_{2}(z_{1},z_{2})}{\partial z_{1}}$
and using (\ref{eq:2ndorderpolefundamental}) with $e^{iQ_{2}(a,z_{2}^{*})}=-1$:

\[
\underset{\mathcal{C}_{J}}{\oint}\frac{\mathrm{d}z_{2}}{2\pi i}\frac{l(a,z_{2})e^{iQ_{2}(a,z_{2})}}{\left[e^{iQ_{2}(z_{1},z_{2})}+1\right]^{2}}=\left.\frac{1}{\frac{\partial Q_{2}(a,z_{2})}{\partial z_{2}}}\frac{\partial}{\partial z_{2}}\left[\frac{l(a,z_{2})}{\frac{\partial Q_{2}(a,z_{2})}{\partial z_{2}}}\right]\right|_{z_{2}=z_{2}^{*}}
\]
Putting it together

\begin{align*}
 & \underset{\mathcal{C}_{a}\times\mathcal{C}_{J}}{\oint\oint}\frac{\mathrm{d}z_{1}}{2\pi i}\frac{\mathrm{d}z_{2}}{2\pi i}\frac{g(z_{1},z_{2})}{\left[a-z_{1}\right]^{2}\left[e^{iQ_{2}(z_{1},z_{2})}+1\right]}\\
 & =\frac{1}{\left.\frac{\partial Q(z_{1},z_{2}^{*})}{\partial z_{1}}\right|_{z_{1}=a}}\left.\left\{ i\left.\frac{\partial g(z_{1},z_{2})}{\partial z_{1}}\right|_{z_{1}=a}-\frac{\partial}{\partial z_{2}}\left[\frac{g(a,z_{2})i\left.\frac{\partial Q(z_{1},z_{2})}{\partial z_{1}}\right|_{z_{1}=a}}{\frac{\partial Q(a,z_{2})}{\partial z_{2}}}\right]\right\} \right|_{z_{2}=z_{2}^{*}}
\end{align*}
where again $z_{2}^{*}$ is the root of 
\[
e^{iQ_{2}(a,z_{2}^{*})}+1=0
\]

\subsection{Useful formulas for practical evaluation}

Using the form of the Bethe-Yang equations for $\vartheta_{1},\vartheta_{2}$
\begin{eqnarray*}
-e^{iQ_{1}(\vartheta_{1},\vartheta_{2})} & = & e^{imL\sinh\vartheta_{1}}S(\vartheta_{1}-\vartheta_{2})=1\\
-e^{iQ_{2}(\vartheta_{1},\vartheta_{2})} & = & e^{imL\sinh\vartheta_{2}}S(\vartheta_{2}-\vartheta_{1})=1
\end{eqnarray*}
we can easily substitute their solutions into the Bethe-Yang equations
for $\vartheta_{1}',\vartheta_{2}'$

\begin{eqnarray*}
e^{iQ_{1}'(\vartheta_{1},\vartheta_{2}')} & = & e^{imL\sinh\vartheta_{1}}\left(-S(\vartheta_{1}-\vartheta_{2}')\right)=-S(\vartheta_{2}-\vartheta_{1})S(\vartheta_{1}-\vartheta_{2}')\\
e^{iQ_{1}'(\vartheta_{2},\vartheta_{2}')} & = & e^{imL\sinh\vartheta_{2}}\left(-S(\vartheta_{2}-\vartheta_{2}')\right)=-S(\vartheta_{1}-\vartheta_{2})S(\vartheta_{2}-\vartheta_{2}')\\
e^{iQ_{2}'(\vartheta_{1}',\vartheta_{1})} & = & e^{imL\sinh\vartheta_{1}}\left(-S(\vartheta_{1}-\vartheta_{1}')\right)=-S(\vartheta_{2}-\vartheta_{1})S(\vartheta_{1}-\vartheta_{1}')\\
e^{iQ_{2}'(\vartheta_{1}',\vartheta_{2})} & = & e^{imL\sinh\vartheta_{2}}\left(-S(\vartheta_{2}-\vartheta_{1}')\right)=-S(\vartheta_{1}-\vartheta_{2})S(\vartheta_{2}-\vartheta_{1}')
\end{eqnarray*}
Denoting 
\[
T\left(\vartheta_{1},\vartheta_{2},\vartheta_{3}\right)=1-S(\vartheta_{1}-\vartheta_{2})S(\vartheta_{2}-\vartheta_{3})
\]
we can also write

\begin{eqnarray*}
e^{iQ_{1}'(\vartheta_{1},\vartheta_{2}')}+1 & = & T\left(\vartheta_{2},\vartheta_{1},\vartheta_{2}'\right)\\
e^{iQ_{1}'(\vartheta_{2},\vartheta_{2}')}+1 & = & T\left(\vartheta_{1},\vartheta_{2},\vartheta_{2}'\right)\\
e^{iQ_{2}'(\vartheta_{1}',\vartheta_{1})}+1 & = & T\left(\vartheta_{2},\vartheta_{1},\vartheta_{1}'\right)\\
e^{iQ_{2}'(\vartheta_{1}',\vartheta_{2})}+1 & = & T\left(\vartheta_{1},\vartheta_{2},\vartheta_{1}'\right)
\end{eqnarray*}

\section{Pole terms for $\tilde{C}_{22}$ \label{sec:Pole-terms-for}}

Here we list the complete result for the pole term subtractions that
appear in

\begin{eqnarray}
\tilde{C}_{22}\left(\vartheta_{1},\vartheta_{2}\right) & = & \frac{1}{2}\underset{\leftrightarrows\times\leftrightarrows}{\oint\oint}\frac{\mathrm{d}\vartheta_{1}'}{2\pi}\frac{\mathrm{d}\vartheta_{2}'}{2\pi}\Bigg[K_{t,x}^{\left(R\right)}\left(\vartheta_{1},\vartheta_{2}|\vartheta_{1}',\vartheta_{2}'\right)\times\nonumber \\
 &  & \frac{F_{4}^{\mathcal{O}_{1}}\left(\vartheta_{2}+i\pi,\vartheta_{1}+i\pi,\vartheta_{1}',\vartheta_{2}'\right)F_{4}^{\mathcal{O}_{2}}\left(\vartheta_{1}+i\pi,\vartheta_{2}+i\pi,\vartheta_{2}',\vartheta_{1}'\right)}{\left[e^{iQ_{1}'\left(\vartheta_{1}',\vartheta_{2}'\right)}+1\right]\left[e^{iQ_{2}'\left(\vartheta_{1}',\vartheta_{2}'\right)}+1\right]}\Bigg]\nonumber \\
 &  & -QF1\left(\vartheta_{1},\vartheta_{2}\right)-QF2\left(\vartheta_{1},\vartheta_{2}\right)-QF3\left(\vartheta_{1},\vartheta_{2}\right)-QF4\left(\vartheta_{1},\vartheta_{2}\right)\nonumber \\
 &  & -QF5\left(\vartheta_{1},\vartheta_{2}\right)-QF6\left(\vartheta_{1},\vartheta_{2}\right)-FF\left(\vartheta_{1},\vartheta_{2}\right)-2SQQ\left(\vartheta_{1},\vartheta_{2}\right)\label{eq:Ctilde22}
\end{eqnarray}
where the $QF$ are the contributions from the $QF$ singularities,
$FF$ comes from the $FF$ singularities and $SQQ$ is the spurious
$QQ$ singularity term. As in the main text, the notation $\leftrightarrows$
corresponds to the straight line contours enclosing the real axis
as illustrated in fig. \ref{fig:Contour-deformation-procedure}. The
explicit form of the individual contributions to (\ref{eq:Ctilde22})
are as follows:

\begin{eqnarray}
QF1\left(\vartheta_{1},\vartheta_{2}\right) & = & \underset{\leftrightarrows}{\oint}\frac{\mathrm{d}\vartheta_{1}'}{2\pi}\left(\frac{F_{4rc}^{\mathcal{O}_{2}}(\vartheta_{1}+i\pi,\vartheta_{2}+i\pi|\vartheta_{1}',\vartheta_{1})F_{2}^{\mathcal{O}_{1}}(\vartheta_{2}+i\pi,\vartheta_{1}')K_{t,x}^{\left(R\right)}(\vartheta_{1},\vartheta_{2}|\vartheta_{1}',\vartheta_{1})}{e^{iQ_{1}'(\vartheta_{1}',\vartheta_{1})}+1}\right.\nonumber \\
 &  & +\left\{ \vartheta_{1}\leftrightarrow\vartheta_{2},\mathcal{O}^{1}\leftrightarrow\mathcal{O}^{2}\right\} \Biggr)\nonumber \\
 & + & \left(\frac{F_{4rc}^{\mathcal{O}_{2}}(\vartheta_{1}+i\pi,\vartheta_{2}+i\pi|\vartheta_{2},\vartheta_{1})F_{2}^{\mathcal{O}_{1}}(i\pi,0)K_{t,x}^{\left(R\right)}(\vartheta_{1},\vartheta_{2},\vartheta_{2},\vartheta_{1})}{mL\cosh\vartheta_{2}+\varphi\left(\vartheta_{2}-\vartheta_{1}\right)}\right.\nonumber \\
 &  & +\left\{ \vartheta_{1}\leftrightarrow\vartheta_{2},\mathcal{O}^{1}\leftrightarrow\mathcal{O}^{2}\right\} \Biggr)\label{eq:QF1}
\end{eqnarray}

\begin{eqnarray}
QF2\left(\vartheta_{1},\vartheta_{2}\right) & = & \underset{\leftrightarrows}{\oint}\frac{\mathrm{d}\vartheta_{1}'}{2\pi}\left(\frac{F_{4rc}^{\mathcal{O}_{1}}(\vartheta_{1}+i\pi,\vartheta_{2}+i\pi|\vartheta_{1}',\vartheta_{1})F_{2}^{\mathcal{O}_{2}}(\vartheta_{2}+i\pi,\vartheta_{1}')K_{t,x}^{\left(R\right)}(\vartheta_{1},\vartheta_{2}|\vartheta_{1}',\vartheta_{1})}{e^{iQ_{1}'(\vartheta_{1}',\vartheta_{1})}+1}\right.\nonumber \\
 &  & +\left\{ \vartheta_{1}\leftrightarrow\vartheta_{2},\mathcal{O}^{1}\leftrightarrow\mathcal{O}^{2}\right\} \Biggr)\nonumber \\
 & + & \left(\frac{F_{4rc}^{\mathcal{O}_{1}}(\vartheta_{1}+i\pi,\vartheta_{2}+i\pi|\vartheta_{2},\vartheta_{1})F_{2}^{\mathcal{O}_{2}}(i\pi,0)K_{t,x}^{\left(R\right)}(\vartheta_{1},\vartheta_{2},\vartheta_{2},\vartheta_{1})}{mL\cosh\vartheta_{2}+\varphi\left(\vartheta_{2}-\vartheta_{1}\right)}\right.\nonumber \\
 &  & +\left\{ \vartheta_{1}\leftrightarrow\vartheta_{2},\mathcal{O}^{1}\leftrightarrow\mathcal{O}^{2}\right\} \Biggr)\label{eq:QF2}
\end{eqnarray}

\begin{eqnarray}
QF3\left(\vartheta_{1},\vartheta_{2}\right) & = & -i\underset{\leftrightarrows}{\oint}\frac{\mathrm{d}\vartheta_{1}'}{2\pi}\left\{ \frac{K_{t,x}^{\left(R\right)}(\vartheta_{1},\vartheta_{2}|\vartheta_{1}',\vartheta_{1})\left[S(\vartheta_{1}-\vartheta_{2})-S(\vartheta_{1}'-\vartheta_{1})\right]}{\left[e^{iQ_{1}'(\vartheta_{1}',\vartheta_{1})}+1\right]\left(\vartheta_{2}-\vartheta_{1}\right)}\right.\nonumber \\
 &  & \left.\times\left[F_{2}^{\mathcal{O}_{1}}(\vartheta_{1}+i\pi,\vartheta_{1}')F_{2}^{\mathcal{O}_{2}}(\vartheta_{2}+i\pi,\vartheta_{1}')+\left\{ \mathcal{O}_{1}\leftrightarrow\mathcal{O}_{2}\right\} \right]+\left\{ \vartheta_{1}\leftrightarrow\vartheta_{2}\right\} \right\} \nonumber \\
 & - & i\left\{ \frac{K_{t,x}^{\left(R\right)}(\vartheta_{1},\vartheta_{2}|\vartheta_{2},\vartheta_{1})\left[S(\vartheta_{1}-\vartheta_{2})-S(\vartheta_{2}-\vartheta_{1})\right]}{\left[mL\cosh\vartheta_{2}+\varphi\left(\vartheta_{2}-\vartheta_{1}\right)\right]\left(\vartheta_{2}-\vartheta_{1}\right)}\right.\nonumber \\
 &  & \left.\times\left[F_{2}^{\mathcal{O}_{1}}(\vartheta_{1}+i\pi,\vartheta_{2})F_{2}^{\mathcal{O}_{2}}(i\pi,0)+\left\{ \mathcal{O}_{1}\leftrightarrow\mathcal{O}_{2}\right\} \right]+\left\{ \vartheta_{1}\leftrightarrow\vartheta_{2}\right\} \right\} \label{eq:QF3}
\end{eqnarray}

\begin{eqnarray}
QF4\left(\vartheta_{1},\vartheta_{2}\right) & = & -i\underset{\leftrightarrows}{\oint}\frac{\mathrm{d}\vartheta_{1}'}{2\pi}\left\{ \frac{K_{t,x}^{\left(R\right)}(\vartheta_{1},\vartheta_{2}|\vartheta_{1}',\vartheta_{1})\left[S(\vartheta_{1}'-\vartheta_{1})S(\vartheta_{1}-\vartheta_{2})-1\right]}{\left[e^{iQ_{1}'(\vartheta_{1}',\vartheta_{1})}+1\right]\left(\vartheta_{2}-\vartheta_{1}'\right)}\right.\nonumber \\
 &  & \left.\times\left[F_{2}^{\mathcal{O}_{1}}(i\pi,0)F_{2}^{\mathcal{O}_{2}}(\vartheta_{2}+i\pi,\vartheta_{1}')+\left\{ \mathcal{O}_{1}\leftrightarrow\mathcal{O}_{2}\right\} \right]+\left\{ \vartheta_{1}\leftrightarrow\vartheta_{2}\right\} \right\} \nonumber \\
 & - & \left\{ \frac{K_{t,x}^{\left(R\right)}(\vartheta_{1},\vartheta_{2}|\vartheta_{2},\vartheta_{1})\varphi\left(\vartheta_{2}-\vartheta_{1}\right)}{\left[mL\cosh\vartheta_{2}+\varphi\left(\vartheta_{2}-\vartheta_{1}\right)\right]}\right.\nonumber \\
 &  & \left.\times\left[F_{2}^{\mathcal{O}_{1}}(i\pi,0)F_{2}^{\mathcal{O}_{2}}(i\pi,0)+\left\{ \mathcal{O}_{1}\leftrightarrow\mathcal{O}_{2}\right\} \right]+\left\{ \vartheta_{1}\leftrightarrow\vartheta_{2}\right\} \right\} \label{eq:QF4}
\end{eqnarray}

\begin{eqnarray}
QF5\left(\vartheta_{1},\vartheta_{2}\right) & = & -i\underset{\leftrightarrows}{\oint}\frac{\mathrm{d}\vartheta_{1}'}{2\pi}\left\{ \frac{K_{t,x}^{\left(R\right)}(\vartheta_{1},\vartheta_{2}|\vartheta_{1}',\vartheta_{1})\left[S(\vartheta_{1}'-\vartheta_{1})-S(\vartheta_{1}-\vartheta_{2})\right]}{\left[e^{iQ_{1}'(\vartheta_{1}',\vartheta_{1})}+1\right]\left(\vartheta_{1}-\vartheta_{1}'\right)}\right.\nonumber \\
 &  & \left.\times\left[F_{2}^{\mathcal{O}_{1}}(\vartheta_{2}+i\pi,\vartheta_{1}')F_{2}^{\mathcal{O}_{2}}(\vartheta_{2}+i\pi,\vartheta_{1})+\left\{ \mathcal{O}_{1}\leftrightarrow\mathcal{O}_{2}\right\} \right]+\left\{ \vartheta_{1}\leftrightarrow\vartheta_{2}\right\} \right\} \nonumber \\
 & - & i\left\{ \frac{K_{t,x}^{\left(R\right)}(\vartheta_{1},\vartheta_{2}|\vartheta_{2},\vartheta_{1})\left[S(\vartheta_{2}-\vartheta_{1})-S(\vartheta_{1}-\vartheta_{2})\right]}{\left[mL\cosh\vartheta_{2}+\varphi\left(\vartheta_{2}-\vartheta_{1}\right)\right]\left(\vartheta_{1}-\vartheta_{2}\right)}\right.\nonumber \\
 &  & \left.\times\left[F_{2}^{\mathcal{O}_{1}}(i\pi,0)F_{2}^{\mathcal{O}_{2}}(\vartheta_{2}+i\pi,\vartheta_{1})+\left\{ \mathcal{O}_{1}\leftrightarrow\mathcal{O}_{2}\right\} \right]+\left\{ \vartheta_{1}\leftrightarrow\vartheta_{2}\right\} \right\} \nonumber \\
 & - & \left\{ K_{t,x}^{\left(R\right)}(\vartheta_{1},\vartheta_{2}|\vartheta_{1},\vartheta_{1})S(\vartheta_{1}-\vartheta_{2})\right.\nonumber \\
 &  & \times\left[F_{2}^{\mathcal{O}}(\vartheta_{2}+i\pi,\vartheta_{1})F_{2}^{\mathcal{O}_{2}}(\vartheta_{2}+i\pi,\vartheta_{1})+\left\{ \mathcal{O}_{1}\leftrightarrow\mathcal{O}_{2}\right\} \right]\nonumber \\
 &  & \left.+\left\{ \vartheta_{1}\leftrightarrow\vartheta_{2}\right\} \right\} \label{eq:QF5}
\end{eqnarray}

\begin{flushleft}
\begin{eqnarray}
QF6\left(\vartheta_{1},\vartheta_{2}\right) & = & -\underset{\leftrightarrows}{\oint}\frac{\mathrm{d}\vartheta_{1}'}{2\pi}\left(\frac{F_{2}^{\mathcal{O}_{1}}(\vartheta_{2}+i\pi,\vartheta_{1}')F_{2}^{\mathcal{O}_{2}}(\vartheta_{2}+i\pi,\vartheta_{1}')K_{t,x}^{\left(R\right)}(\vartheta_{1},\vartheta_{2}|\vartheta_{1}',\vartheta_{1})}{\left[e^{iQ_{1}'(\vartheta_{1}',\vartheta_{1})}+1\right]}\right.\nonumber \\
 &  & \times\left\{ T\left(\vartheta_{1}',\vartheta_{1},\vartheta_{2}\right)\left[-mx\cosh\vartheta_{1}+imt\sinh\vartheta_{1}\right]-mL\cosh\left(\vartheta_{1}\right)\right.\nonumber \\
 &  & \left.+\varphi(\vartheta_{1}'-\vartheta_{1})S(\vartheta_{1}'-\vartheta_{1})S(\vartheta_{1}-\vartheta_{2})\right\} +\left\{ \vartheta_{1}\leftrightarrow\vartheta_{2}\right\} \Biggr)\nonumber \\
 & + & F_{2}^{\mathcal{O}_{1}}(i\pi,0)F_{2}^{\mathcal{O}_{2}}(i\pi,0)K_{t,x}^{\left(R\right)}(\vartheta_{1},\vartheta_{2}|\vartheta_{2},\vartheta_{1})\nonumber \\
 &  & \times\left(\frac{\left[mL\cosh\vartheta_{1}-\varphi\left(\vartheta_{1}-\vartheta_{2}\right)\right]}{\left[mL\cosh\vartheta_{2}+\varphi\left(\vartheta_{2}-\vartheta_{1}\right)\right]}+\frac{\left[mL\cosh\vartheta_{2}-\varphi\left(\vartheta_{2}-\vartheta_{1}\right)\right]}{\left[mL\cosh\vartheta_{1}+\varphi\left(\vartheta_{1}-\vartheta_{2}\right)\right]}\right)\nonumber \\
 & + & \underset{\leftrightarrows}{\oint}\frac{\mathrm{d}\vartheta_{1}'}{2\pi}\Biggl(\frac{i}{\left[e^{iQ_{1}'(\vartheta_{1}',\vartheta_{1})}+1\right]}\nonumber \\
 &  & \times\frac{\partial}{\partial\vartheta_{1}'}\left[\frac{K_{t,x}^{\left(R\right)}(\vartheta_{1},\vartheta_{2}|\vartheta_{1}',\vartheta_{1})T\left(\vartheta_{1}',\vartheta_{1},\vartheta_{2}\right)\varphi\left(\vartheta_{1}'-\vartheta_{1}\right)}{\left[mL\cosh\vartheta_{1}'+\varphi\left(\vartheta_{1}'-\vartheta_{1}\right)\right]}\right.\nonumber \\
 &  & \times F_{2}^{\mathcal{O}_{2}}(\vartheta_{2}+i\pi,\vartheta_{1}')F_{2}^{\mathcal{O}_{1}}(\vartheta_{2}+i\pi,\vartheta_{1}')\Biggr]+\left\{ \vartheta_{1}\leftrightarrow\vartheta_{2}\right\} \Biggr)\nonumber \\
 & + & K_{t,x}^{\left(R\right)}(\vartheta_{1},\vartheta_{2}|\vartheta_{2},\vartheta_{1})F_{2}^{\mathcal{O}_{1}}(i\pi,0)F_{2}^{\mathcal{O}_{2}}(i\pi,0)\varphi\left(\vartheta_{2}-\vartheta_{1}\right)^{2}\label{eq:QF6}\\
 &  & \times\left(\frac{1}{\left[mL\cosh\vartheta_{2}+\varphi\left(\vartheta_{2}-\vartheta_{1}\right)\right]^{2}}+\frac{1}{\left[mL\cosh\vartheta_{1}+\varphi\left(\vartheta_{1}-\vartheta_{2}\right)\right]^{2}}\right)\nonumber 
\end{eqnarray}

\par\end{flushleft}

\begin{eqnarray}
FF & = & K_{t,x}^{\left(R\right)}\left(\vartheta_{1},\vartheta_{2},\vartheta_{1},\vartheta_{1}\right)S\left(\vartheta_{1}-\vartheta_{2}\right)F_{2}^{\mathcal{O}_{1}}\left(\vartheta_{2}+i\pi,\vartheta_{1}\right)F_{2}^{\mathcal{O}_{2}}\left(\vartheta_{2}+i\pi,\vartheta_{1}\right)\nonumber \\
 &  & +\left\{ \vartheta_{1}\leftrightarrow\vartheta_{2}\right\} \label{eq:FF}
\end{eqnarray}
 
\begin{eqnarray}
SQQ\left(\vartheta_{1},\vartheta_{2}\right) & = & \underset{\mathcal{C}_{\vartheta_{1}}\times\mathcal{C}_{\vartheta_{2}}}{\oint\oint}\frac{\mathrm{d}\vartheta_{1}'}{2\pi}\frac{\mathrm{d}\vartheta_{2}'}{2\pi}\Bigg\{\frac{F_{4}^{\mathcal{O}_{1}}\left(\vartheta_{2}+i\pi,\vartheta_{1}+i\pi,\vartheta_{1}',\vartheta_{2}'\right)F_{4}^{\mathcal{O}_{2}}\left(\vartheta_{1}+i\pi,\vartheta_{2}+i\pi,\vartheta_{2}',\vartheta_{1}'\right)}{\left[e^{iQ_{1}'\left(\vartheta_{1}',\vartheta_{2}'\right)}+1\right]\left[e^{iQ_{2}'\left(\vartheta_{1}',\vartheta_{2}'\right)}+1\right]}\nonumber \\
 &  & \times K_{t,x}^{\left(R\right)}\left(\vartheta_{1},\vartheta_{2}|\vartheta_{1}',\vartheta_{2}'\right)\Bigg\}\label{eq:SQQ}
\end{eqnarray}
We gave these contributions in their exact finite volume form (i.e.
including the full volume dependence): albeit they simplify when taking
the volume to infinity, and the $SQQ$ term does not even contribute
in this limit, all terms must be kept in order for the numerical verification
of Section \ref{sec:Numerical-verification-of} to work properly.

\section{Symmetry of $D_{22}$\label{sec:Symmetry-of-D22}}

We want to prove that $D_{22}$ is symmetric under

\begin{eqnarray}
t & \to & R-t\nonumber \\
\mathcal{O}^{1} & \leftrightarrow & \mathcal{O}^{2}\nonumber \\
x & \to & -x\label{eq:d22symmetry}
\end{eqnarray}
First of all, notice that the diagonal terms contain no $t$ and $x$
factor, and are manifestly symmetric under exchanging $\mathcal{O}_{1}$
and $\mathcal{O}_{2}$ . So it remains only to treat the non-diagonal
part.

\subsection{The four-integral term}

First we treat the term in (\ref{eq:d22final}) that contains a fourfold
integral. After the transformation we change the variables $\vartheta_{1,2}\leftrightarrow\vartheta_{1,2}'$,
and shift every contour with $-2i\varepsilon$. This results in the
contour now running under the real axis:

\begin{eqnarray*}
 &  & \frac{1}{4}\iiiint\frac{\mathrm{d}\vartheta_{1}}{2\pi}\frac{\mathrm{d}\vartheta_{2}}{2\pi}\frac{\mathrm{d}\vartheta_{1}'}{2\pi}\frac{\mathrm{d}\vartheta_{2}'}{2\pi}K_{t,x}^{\left(R\right)}(\vartheta_{1},\vartheta_{2}|\vartheta_{1}'-i\varepsilon,\vartheta_{2}'-i\varepsilon)\\
 &  & \times F_{4}^{\mathcal{O}_{1}}(\vartheta_{2}+i\pi,\vartheta_{1}+i\pi,\vartheta_{1}'-i\varepsilon,\vartheta_{2}'-i\varepsilon)F_{4}^{\mathcal{O}_{2}}(\vartheta_{1}+i\pi,\vartheta_{2}+i\pi,\vartheta_{2}'-i\varepsilon,\vartheta_{1}'-i\varepsilon)
\end{eqnarray*}
By shifting the contour above the real axis we can transform this
term back to its form in (\ref{eq:d22final}), but during this process
we pick up some pole contributions from the poles in eqn. (\ref{eq:F4F4poleterms}).

\subsubsection{First order pole terms containing $F_{4rc}$}

Using eqn. (\ref{eq:F4F4poleterms}) we can identify a contribution
containing $F_{4rc}$. In these contribution, all poles are of first
order, so one can apply the Cauchy formula directly. One set of such
terms is given by

\begin{eqnarray*}
 &  & \frac{1}{4}\iint_{\mathcal{C}_{-}}\frac{\mathrm{d}\vartheta_{1}'}{2\pi}\frac{\mathrm{d}\vartheta_{2}'}{2\pi}F_{4rc}^{\mathcal{O}_{1}}(\vartheta_{2}+i\pi,\vartheta_{1}+i\pi|\vartheta_{1}',\vartheta_{2}')\left[\frac{E}{\vartheta_{1}-\vartheta_{2}'}+\frac{F}{\vartheta_{1}-\vartheta_{1}'}+\frac{G}{\vartheta_{2}-\vartheta_{2}'}+\frac{H}{\vartheta_{2}-\vartheta_{1}'}\right]\\
 &  & \times K_{t,x}^{\left(R\right)}(\vartheta_{1},\vartheta_{2}|\vartheta_{1}',\vartheta_{2}')\\
 & = & -\frac{1}{4}\iint_{\mathcal{C}_{+}}\frac{\mathrm{d}\vartheta_{1}'}{2\pi}\frac{\mathrm{d}\vartheta_{2}'}{2\pi}F_{4rc}^{\mathcal{O}_{1}}(\vartheta_{2}+i\pi,\vartheta_{1}+i\pi|\vartheta_{1}',\vartheta_{2}')\left[\frac{E}{\vartheta_{2}'-\vartheta_{1}}+\frac{F}{\vartheta_{1}'-\vartheta_{1}}+\frac{G}{\vartheta_{2}'-\vartheta_{2}}+\frac{H}{\vartheta_{1}'-\vartheta_{2}}\right]\\
 &  & \times K_{t,x}^{\left(R\right)}(\vartheta_{1},\vartheta_{2}|\vartheta_{1}',\vartheta_{2}')\\
 & + & \frac{1}{4}\int\frac{\mathrm{d}\vartheta_{1}'}{2\pi}F_{4rc}^{\mathcal{O}_{1}}(\vartheta_{2}+i\pi,\vartheta_{1}+i\pi|\vartheta_{1}',\vartheta_{1})F_{2}^{\mathcal{O}_{2}}(\vartheta_{2}+i\pi,\vartheta_{1}')K_{t,x}^{\left(R\right)}(\vartheta_{1},\vartheta_{2}|\vartheta_{1}',\vartheta_{1})\\
 &  & \times\left[S(\vartheta_{1}-\vartheta_{2})-S(\vartheta_{1}-\vartheta_{1}')\right]\\
 & + & \frac{1}{4}\int\frac{\mathrm{d}\vartheta_{2}'}{2\pi}F_{4rc}^{\mathcal{O}_{1}}(\vartheta_{2}+i\pi,\vartheta_{1}+i\pi|\vartheta_{1},\vartheta_{2}')F_{2}^{\mathcal{O}_{2}}(\vartheta_{2}+i\pi,\vartheta_{2}')K_{t,x}^{\left(R\right)}(\vartheta_{1},\vartheta_{2}|\vartheta_{1},\vartheta_{2}')\\
 &  & \times\left[S(\vartheta_{2}'-\vartheta_{1})S(\vartheta_{1}-\vartheta_{2})-1\right]\\
 & + & \frac{1}{4}\int\frac{\mathrm{d}\vartheta_{1}'}{2\pi}F_{4rc}^{\mathcal{O}_{1}}(\vartheta_{2}+i\pi,\vartheta_{1}+i\pi|\vartheta_{1}',\vartheta_{2})F_{2}^{\mathcal{O}_{2}}(\vartheta_{1}+i\pi,\vartheta_{1}')K_{t,x}^{\left(R\right)}(\vartheta_{1},\vartheta_{2}|\vartheta_{1}',\vartheta_{2})\\
 &  & \times\left[1-S(\vartheta_{1}-\vartheta_{2})S(\vartheta_{2}-\vartheta_{1}')\right]\\
 & + & \frac{1}{4}\int\frac{\mathrm{d}\vartheta_{2}'}{2\pi}F_{4rc}^{\mathcal{O}_{1}}(\vartheta_{2}+i\pi,\vartheta_{1}+i\pi|\vartheta_{2},\vartheta_{2}')F_{2}^{\mathcal{O}_{2}}(\vartheta_{1}+i\pi,\vartheta_{2}')K_{t,x}^{\left(R\right)}(\vartheta_{1},\vartheta_{2}|\vartheta_{2},\vartheta_{2}')\\
 &  & \times\left[S(\vartheta_{2}'-\vartheta_{2})-S(\vartheta_{1}-\vartheta_{2})\right]
\end{eqnarray*}
and a similar contribution from

\begin{eqnarray*}
 &  & \frac{1}{4}\iint_{\mathcal{C}_{-}}\frac{\mathrm{d}\vartheta_{1}'}{2\pi}\frac{\mathrm{d}\vartheta_{2}'}{2\pi}F_{4rc}^{\mathcal{O}_{2}}(\vartheta_{1}+i\pi,\vartheta_{2}+i\pi|\vartheta_{2}',\vartheta_{1}')\left[\frac{A}{\vartheta_{2}-\vartheta_{1}'}+\frac{B}{\vartheta_{2}-\vartheta_{2}'}+\frac{C}{\vartheta_{1}-\vartheta_{1}'}+\frac{D}{\vartheta_{1}-\vartheta_{2}'}\right]\\
 &  & \times K_{t,x}^{\left(R\right)}(\vartheta_{1},\vartheta_{2}|\vartheta_{1}',\vartheta_{2}')\\
 & = & -\frac{1}{4}\iint_{\mathcal{C}_{+}}\frac{\mathrm{d}\vartheta_{1}'}{2\pi}\frac{\mathrm{d}\vartheta_{2}'}{2\pi}F_{4rc}^{\mathcal{O}_{2}}(\vartheta_{1}+i\pi,\vartheta_{2}+i\pi|\vartheta_{2}',\vartheta_{1}')\left[\frac{A}{\vartheta_{1}'-\vartheta_{2}}+\frac{B}{\vartheta_{2}'-\vartheta_{2}}+\frac{C}{\vartheta_{1}'-\vartheta_{1}}+\frac{D}{\vartheta_{2}'-\vartheta_{1}}\right]\\
 &  & \times K_{t,x}^{\left(R\right)}(\vartheta_{1},\vartheta_{2}|\vartheta_{1}',\vartheta_{2}')\\
 & + & \frac{1}{4}\int\frac{\mathrm{d}\vartheta_{2}'}{2\pi}F_{4rc}^{\mathcal{O}_{2}}(\vartheta_{1}+i\pi,\vartheta_{2}+i\pi|\vartheta_{2}',\vartheta_{2})F_{2}^{\mathcal{O}_{1}}(\vartheta_{1}+i\pi,\vartheta_{2}')K_{t,x}^{\left(R\right)}(\vartheta_{1},\vartheta_{2}|\vartheta_{2},\vartheta_{2}')\\
 &  & \times\left[S(\vartheta_{2}-\vartheta_{1})-S(\vartheta_{2}-\vartheta_{2}')\right]\\
 & + & \frac{1}{4}\int\frac{\mathrm{d}\vartheta_{1}'}{2\pi}F_{4rc}^{\mathcal{O}_{2}}(\vartheta_{1}+i\pi,\vartheta_{2}+i\pi|\vartheta_{2},\vartheta_{1}')F_{2}^{\mathcal{O}_{1}}(\vartheta_{1}+i\pi,\vartheta_{1}')K_{t,x}^{\left(R\right)}(\vartheta_{1},\vartheta_{2}|\vartheta_{1}',\vartheta_{2})\\
 &  & \times\left[S(\vartheta_{1}'-\vartheta_{2})S(\vartheta_{2}-\vartheta_{1})-1\right]\\
 & + & \frac{1}{4}\int\frac{\mathrm{d}\vartheta_{2}'}{2\pi}F_{4rc}^{\mathcal{O}_{2}}(\vartheta_{1}+i\pi,\vartheta_{2}+i\pi|\vartheta_{2}',\vartheta_{1})F_{2}^{\mathcal{O}_{1}}(\vartheta_{2}+i\pi,\vartheta_{2}')K_{t,x}^{\left(R\right)}(\vartheta_{1},\vartheta_{2}|\vartheta_{1},\vartheta_{2}')\\
 &  & \times\left[1-S(\vartheta_{2}-\vartheta_{1})S(\vartheta_{1}-\vartheta_{2}')\right]\\
 & + & \frac{1}{4}\int\frac{\mathrm{d}\vartheta_{1}'}{2\pi}F_{4rc}^{\mathcal{O}_{2}}(\vartheta_{1}+i\pi,\vartheta_{2}+i\pi|\vartheta_{1},\vartheta_{1}')F_{2}^{\mathcal{O}_{1}}(\vartheta_{2}+i\pi,\vartheta_{1}')K_{t,x}^{\left(R\right)}(\vartheta_{1},\vartheta_{2}|\vartheta_{1}',\vartheta_{1})\\
 &  & \times\left[S(\vartheta_{1}'-\vartheta_{1})-S(\vartheta_{2}-\vartheta_{1})\right]
\end{eqnarray*}
where $\mathcal{C}_{\pm}$ denote integration running above/below
from the real axis, in the direction from left to right.

Since the exchange property (\ref{eq:exchangeaxiom}) is valid for
$F_{4rc}$ in the first and last two variables, it can be used to
simplify the total contribution to

\begin{eqnarray*}
 &  & \frac{1}{2}\int\frac{\mathrm{d}\vartheta_{1}'}{2\pi}\Bigg\{ K_{t,x}^{\left(R\right)}(\vartheta_{1},\vartheta_{2}|\vartheta_{1}',\vartheta_{1})\left\{ F_{4rc}^{\mathcal{O}_{1}}(\vartheta_{1}+i\pi,\vartheta_{2}+i\pi|\vartheta_{1}',\vartheta_{1})F_{2}^{\mathcal{O}_{2}}(\vartheta_{2}+i\pi,\vartheta_{1}')+\left\{ \mathcal{O}^{1}\leftrightarrow\mathcal{O}^{2}\right\} \right\} \\
 &  & +\left\{ \vartheta_{1}\leftrightarrow\vartheta_{2}\right\} \Bigg\}\\
 & - & \frac{1}{2}\int\frac{\mathrm{d}\vartheta_{1}'}{2\pi}\Bigg\{ K_{t,x}^{\left(R\right)}(\vartheta_{1},\vartheta_{2}|\vartheta_{1}',\vartheta_{1})\left\{ F_{4rc}^{\mathcal{O}_{1}}(\vartheta_{2}+i\pi,\vartheta_{1}+i\pi|\vartheta_{1},\vartheta_{1}')F_{2}^{\mathcal{O}_{2}}(\vartheta_{2}+i\pi,\vartheta_{1}')+\left\{ \mathcal{O}^{1}\leftrightarrow\mathcal{O}^{2}\right\} \right\} \\
 &  & +\left\{ \vartheta_{1}\leftrightarrow\vartheta_{2}\right\} \Bigg\}
\end{eqnarray*}

\subsubsection{First order poles without $F_{4rc}$}

The form of these terms is 
\begin{eqnarray*}
 &  & \frac{1}{4}\iiiint_{\mathcal{C}_{-}}\frac{\mathrm{d}\vartheta_{1}}{2\pi}\frac{\mathrm{d}\vartheta_{2}}{2\pi}\frac{\mathrm{d}\vartheta_{1}'}{2\pi}\frac{\mathrm{d}\vartheta_{2}'}{2\pi}K_{t,x}^{\left(R\right)}(\vartheta_{1},\vartheta_{2}|\vartheta_{1}',\vartheta_{2}')\times\\
 & \times & \left[\frac{AE+DH}{\left(\vartheta_{2}-\vartheta_{1}'\right)\left(\vartheta_{1}-\vartheta_{2}'\right)}+\frac{AF+CH}{\left(\vartheta_{2}-\vartheta_{1}'\right)\left(\vartheta_{1}-\vartheta_{1}'\right)}+\frac{AG+BH}{\left(\vartheta_{2}-\vartheta_{1}'\right)\left(\vartheta_{2}-\vartheta_{2}'\right)}\right.\\
 &  & \left.+\frac{BE+DG}{\left(\vartheta_{2}-\vartheta_{2}'\right)\left(\vartheta_{1}-\vartheta_{2}'\right)}+\frac{BF+CG}{\left(\vartheta_{2}-\vartheta_{2}'\right)\left(\vartheta_{1}-\vartheta_{1}'\right)}+\frac{CE+DF}{\left(\vartheta_{1}-\vartheta_{1}'\right)\left(\vartheta_{1}-\vartheta_{2}'\right)}\right]
\end{eqnarray*}
Their evaluation is tedious, but straightforward. For example

\begin{eqnarray*}
AE+DH & = & -\left(S(\vartheta_{2}-\vartheta_{1})-S(\vartheta_{1}'-\vartheta_{2}')\right)\left(S(\vartheta_{1}-\vartheta_{2})-S(\vartheta_{2}'-\vartheta_{1}')\right)\\
 &  & \times\left\{ F_{2}^{\mathcal{O}_{1}}(\vartheta_{1}+i\pi,\vartheta_{2}')F_{2}^{\mathcal{O}_{2}}(\vartheta_{2}+i\pi,\vartheta_{1}')+\left\{ \mathcal{O}^{1}\leftrightarrow\mathcal{O}^{2}\right\} \right\} 
\end{eqnarray*}
and one can evaluate the contributions resulting from the contour
shift as 
\begin{eqnarray*}
 &  & +\frac{1}{4}\iiiint_{\mathcal{C}_{-}}\frac{\mathrm{d}\vartheta_{1}}{2\pi}\frac{\mathrm{d}\vartheta_{2}}{2\pi}\frac{\mathrm{d}\vartheta_{1}'}{2\pi}\frac{\mathrm{d}\vartheta_{2}'}{2\pi}K_{t,x}^{\left(R\right)}(\vartheta_{1},\vartheta_{2}|\vartheta_{1}',\vartheta_{2}')\frac{AE+DH}{\left(\vartheta_{1}'-\vartheta_{2}\right)\left(\vartheta_{2}'-\vartheta_{1}\right)}\\
 & = & +\frac{1}{4}\iiiint_{\mathcal{C}_{+}}\frac{\mathrm{d}\vartheta_{1}}{2\pi}\frac{\mathrm{d}\vartheta_{2}}{2\pi}\frac{\mathrm{d}\vartheta_{1}'}{2\pi}\frac{\mathrm{d}\vartheta_{2}'}{2\pi}K_{t,x}^{\left(R\right)}(\vartheta_{1},\vartheta_{2}|\vartheta_{1}',\vartheta_{2}')\frac{AE+DH}{\left(\vartheta_{1}'-\vartheta_{2}\right)\left(\vartheta_{2}'-\vartheta_{1}\right)}\\
 &  & +\frac{1}{4}\iiint_{\mathcal{C}_{+}}\frac{\mathrm{d}\vartheta_{1}}{2\pi}\frac{\mathrm{d}\vartheta_{2}}{2\pi}\frac{\mathrm{d}\vartheta_{1}'}{2\pi}\frac{K_{t,x}^{\left(R\right)}(\vartheta_{1},\vartheta_{2}|\vartheta_{1}',\vartheta_{1})}{\left(\vartheta_{1}'-\vartheta_{2}\right)}i\left.\left[AE+DH\right]\right|_{\vartheta_{2}'=\vartheta_{1}}\\
 &  & +\frac{1}{4}\iiint_{\mathcal{C}_{+}}\frac{\mathrm{d}\vartheta_{1}}{2\pi}\frac{\mathrm{d}\vartheta_{2}}{2\pi}\frac{\mathrm{d}\vartheta_{2}'}{2\pi}\frac{K_{t,x}^{\left(R\right)}(\vartheta_{1},\vartheta_{2}|\vartheta_{2},\vartheta_{2}')}{\left(\vartheta_{2}'-\vartheta_{1}\right)}i\left.\left[AE+DH\right]\right|_{\vartheta_{1}'=\vartheta_{2}}\\
 &  & -\frac{1}{4}\iiint_{\mathcal{C}_{+}}\frac{\mathrm{d}\vartheta_{1}}{2\pi}\frac{\mathrm{d}\vartheta_{2}}{2\pi}\frac{\mathrm{d}\vartheta_{2}'}{2\pi}K_{t,x}^{\left(R\right)}(\vartheta_{1},\vartheta_{2}|\vartheta_{2},\vartheta_{1})\left.\left[AE+DH\right]\right|_{\vartheta_{1}'=\vartheta_{2},\:\vartheta_{2}'=\vartheta_{1}}\\
 & = & +\frac{1}{4}\iiiint_{\mathcal{C}_{+}}\frac{\mathrm{d}\vartheta_{1}}{2\pi}\frac{\mathrm{d}\vartheta_{2}}{2\pi}\frac{\mathrm{d}\vartheta_{1}'}{2\pi}\frac{\mathrm{d}\vartheta_{2}'}{2\pi}K_{t,x}^{\left(R\right)}(\vartheta_{1},\vartheta_{2}|\vartheta_{1}',\vartheta_{2}')\frac{AE+DH}{\left(\vartheta_{1}'-\vartheta_{2}\right)\left(\vartheta_{2}'-\vartheta_{1}\right)}\\
 &  & -\frac{i}{2}\iint\frac{\mathrm{d}\vartheta_{1}}{2\pi}\frac{\mathrm{d}\vartheta_{2}}{2\pi}\wp\int\frac{\mathrm{d}\vartheta_{1}'}{2\pi}\frac{K_{t,x}^{\left(R\right)}(\vartheta_{1},\vartheta_{2}|\vartheta_{1}',\vartheta_{1})}{\left(\vartheta_{1}'-\vartheta_{2}\right)}\left\{ F_{2}^{\mathcal{O}_{1}}(i\pi,0)F_{2}^{\mathcal{O}_{2}}(\vartheta_{2}+i\pi,\vartheta_{1}')+\left\{ \mathcal{O}^{1}\leftrightarrow\mathcal{O}^{2}\right\} \right\} \\
 &  & \times\left[S(\vartheta_{2}-\vartheta_{1})-S(\vartheta_{1}'-\vartheta_{1})\right]\left[S(\vartheta_{1}-\vartheta_{2})-S(\vartheta_{1}-\vartheta_{1}')\right]
\end{eqnarray*}
and similarly for the remaining five cases.

\subsubsection{Second order poles}

We get four contributions which contain a second order pole. All four
are the same after relabeling the rapidities:

\begin{eqnarray*}
 &  & +\frac{1}{4}\iiiint_{\mathcal{C}_{-}}\frac{\mathrm{d}\vartheta_{1}}{2\pi}\frac{\mathrm{d}\vartheta_{2}}{2\pi}\frac{\mathrm{d}\vartheta_{1}'}{2\pi}\frac{\mathrm{d}\vartheta_{2}'}{2\pi}\Bigg\{\left[\frac{AH}{\left(\vartheta_{2}-\vartheta_{1}'\right)^{2}}+\frac{BG}{\left(\vartheta_{2}-\vartheta_{2}'\right)^{2}}+\frac{CF}{\left(\vartheta_{1}-\vartheta_{1}'\right)^{2}}+\frac{DE}{\left(\vartheta_{1}-\vartheta_{2}'\right)^{2}}\right]\\
 &  & \times K_{t,x}^{\left(R\right)}(\vartheta_{1},\vartheta_{2}|\vartheta_{1}',\vartheta_{2}')\Bigg\}\\
 & = & \iiiint_{\mathcal{C}_{-}}\frac{\mathrm{d}\vartheta_{1}}{2\pi}\frac{\mathrm{d}\vartheta_{2}}{2\pi}\frac{\mathrm{d}\vartheta_{1}'}{2\pi}\frac{\mathrm{d}\vartheta_{2}'}{2\pi}\frac{DE}{\left(\vartheta_{1}-\vartheta_{2}'\right)^{2}}K_{t,x}^{\left(R\right)}(\vartheta_{1},\vartheta_{2}|\vartheta_{1}',\vartheta_{2}')\\
 & = & \iiiint_{\mathcal{C}_{+}}\frac{\mathrm{d}\vartheta_{1}}{2\pi}\frac{\mathrm{d}\vartheta_{2}}{2\pi}\frac{\mathrm{d}\vartheta_{1}'}{2\pi}\frac{\mathrm{d}\vartheta_{2}'}{2\pi}\frac{DE}{\left(\vartheta_{1}-\vartheta_{2}'\right)^{2}}K_{t,x}^{\left(R\right)}(\vartheta_{1},\vartheta_{2}|\vartheta_{1}',\vartheta_{2}')\\
 &  & +i\iiiint\frac{\mathrm{d}\vartheta_{1}}{2\pi}\frac{\mathrm{d}\vartheta_{2}}{2\pi}\frac{\mathrm{d}\vartheta_{1}'}{2\pi}\left.\frac{\partial\left[DEK_{t,x}^{\left(R\right)}(\vartheta_{1},\vartheta_{2}|\vartheta_{1}',\vartheta_{2}')\right]}{\partial\vartheta_{2}'}\right|_{\vartheta_{2}'=\vartheta_{1}}
\end{eqnarray*}
After performing the differentiation, the final result is:

\begin{eqnarray*}
 &  & \iiiint_{\mathcal{C}_{+}}\frac{\mathrm{d}\vartheta_{1}}{2\pi}\frac{\mathrm{d}\vartheta_{2}}{2\pi}\frac{\mathrm{d}\vartheta_{1}'}{2\pi}\frac{\mathrm{d}\vartheta_{2}'}{2\pi}\frac{DE}{\left(\vartheta_{1}-\vartheta_{2}'\right)^{2}}K_{t,x}^{\left(R\right)}(\vartheta_{1},\vartheta_{2}|\vartheta_{1}',\vartheta_{2}')\\
 &  & +\iiint\frac{\mathrm{d}\vartheta_{1}}{2\pi}\frac{\mathrm{d}\vartheta_{2}}{2\pi}\frac{\mathrm{d}\vartheta_{1}'}{2\pi}F_{2}^{\mathcal{O}_{1}}(\vartheta_{2}+i\pi,\vartheta_{1}')F_{2}^{\mathcal{O}_{2}}(\vartheta_{2}+i\pi,\vartheta_{1}')K_{t,x}^{\left(R\right)}(\vartheta_{1},\vartheta_{2}|\vartheta_{1}',\vartheta_{1})\\
 &  & \times\left\{ \left[mx\cosh\vartheta_{1}-imt\sinh\vartheta_{1}\right]T\left(\vartheta_{2},\vartheta_{1},\vartheta_{1}'\right)T\left(\vartheta_{1}',\vartheta_{1},\vartheta_{2}\right)\right.\\
 &  & \left.+\left[S(\vartheta_{2}-\vartheta_{1})S(\vartheta_{1}-\vartheta_{1}')-S(\vartheta_{1}'-\vartheta_{1})S(\vartheta_{1}-\vartheta_{2})\right]\varphi\left(\vartheta_{1}-\vartheta_{1}'\right)\right\} 
\end{eqnarray*}
where the notation 
\[
T\left(\vartheta_{1},\vartheta_{2},\vartheta_{3}\right)=1-S(\vartheta_{1}-\vartheta_{2})S(\vartheta_{2}-\vartheta_{3})
\]
was used.

\subsubsection{Putting together the results}

Putting together the result for the fourfold integral term one obtains

\begin{eqnarray*}
 &  & \frac{1}{4}\iiiint\frac{\mathrm{d}\vartheta_{1}}{2\pi}\frac{\mathrm{d}\vartheta_{2}}{2\pi}\frac{\mathrm{d}\vartheta_{1}'}{2\pi}\frac{\mathrm{d}\vartheta_{2}'}{2\pi}F_{4}^{\mathcal{O}_{1}}(\vartheta_{2}+i\pi,\vartheta_{1}+i\pi,\vartheta_{1}'-i\varepsilon,\vartheta_{2}'-i\varepsilon)\\
 &  & \times F_{4}^{\mathcal{O}_{2}}(\vartheta_{1}+i\pi,\vartheta_{2}+i\pi,\vartheta_{2}'-i\varepsilon,\vartheta_{1}'-i\varepsilon)K_{t,x}^{\left(R\right)}(\vartheta_{1},\vartheta_{2}|\vartheta_{1}'-i\varepsilon,\vartheta_{2}'-i\varepsilon)\\
 & = & \frac{1}{4}\iiiint\frac{\mathrm{d}\vartheta_{1}}{2\pi}\frac{\mathrm{d}\vartheta_{2}}{2\pi}\frac{\mathrm{d}\vartheta_{1}'}{2\pi}\frac{\mathrm{d}\vartheta_{2}'}{2\pi}F_{4}^{\mathcal{O}_{1}}(\vartheta_{2}+i\pi,\vartheta_{1}+i\pi,\vartheta_{1}'+i\varepsilon,\vartheta_{2}'+i\varepsilon)\\
 &  & \times F_{4}^{\mathcal{O}_{2}}(\vartheta_{1}+i\pi,\vartheta_{2}+i\pi,\vartheta_{2}'+i\varepsilon,\vartheta_{1}'+i\varepsilon)K_{t,x}^{\left(R\right)}(\vartheta_{1},\vartheta_{2}|\vartheta_{1}'+i\varepsilon,\vartheta_{2}'+i\varepsilon)\\
 & + & \iiint\frac{\mathrm{d}\vartheta_{1}}{2\pi}\frac{\mathrm{d}\vartheta_{2}}{2\pi}\frac{\mathrm{d}\vartheta_{1}'}{2\pi}K_{t,x}^{\left(R\right)}(\vartheta_{1},\vartheta_{2}|\vartheta_{1}',\vartheta_{1})\Big\{ F_{4rc}^{\mathcal{O}_{1}}(\vartheta_{1}+i\pi,\vartheta_{2}+i\pi|\vartheta_{1}',\vartheta_{1})F_{2}^{\mathcal{O}_{2}}(\vartheta_{2}+i\pi,\vartheta_{1}')\\
 &  & +\left\{ \mathcal{O}^{1}\leftrightarrow\mathcal{O}^{2}\right\} \Big\}\\
 & - & \iiint\frac{\mathrm{d}\vartheta_{1}}{2\pi}\frac{\mathrm{d}\vartheta_{2}}{2\pi}\frac{\mathrm{d}\vartheta_{1}'}{2\pi}K_{t,x}^{\left(R\right)}(\vartheta_{1},\vartheta_{2}|\vartheta_{1}',\vartheta_{1})\Big\{ F_{4rc}^{\mathcal{O}_{1}}(\vartheta_{2}+i\pi,\vartheta_{1}+i\pi|\vartheta_{1},\vartheta_{1}')F_{2}^{\mathcal{O}_{2}}(\vartheta_{2}+i\pi,\vartheta_{1}')\\
 &  & +\left\{ \mathcal{O}^{1}\leftrightarrow\mathcal{O}^{2}\right\} \Big\}\\
 & - & i\iint\frac{\mathrm{d}\vartheta_{1}}{2\pi}\frac{\mathrm{d}\vartheta_{2}}{2\pi}\wp\int\frac{\mathrm{d}\vartheta_{1}'}{2\pi}\frac{K_{t,x}^{\left(R\right)}(\vartheta_{1},\vartheta_{2}|\vartheta_{1}',\vartheta_{1})}{\left(\vartheta_{2}-\vartheta_{1}'\right)}\Big\{ F_{2}^{\mathcal{O}_{1}}(i\pi,0)F_{2}^{\mathcal{O}_{2}}(\vartheta_{2}+i\pi,\vartheta_{1}')\\
 &  & +\left\{ \mathcal{O}^{1}\leftrightarrow\mathcal{O}^{2}\right\} \Big\}\left[S(\vartheta_{1}'-\vartheta_{1})-S(\vartheta_{2}-\vartheta_{1})\right]\left[S(\vartheta_{1}-\vartheta_{2})-S(\vartheta_{1}-\vartheta_{1}')\right]\\
 & - & i\iint\frac{\mathrm{d}\vartheta_{1}}{2\pi}\frac{\mathrm{d}\vartheta_{2}}{2\pi}\wp\int\frac{\mathrm{d}\vartheta_{1}'}{2\pi}\frac{K_{t,x}^{\left(R\right)}(\vartheta_{1},\vartheta_{2}|\vartheta_{1}',\vartheta_{1})}{\left(\vartheta_{1}-\vartheta_{1}'\right)}\Big\{ F_{2}^{\mathcal{O}_{1}}(\vartheta_{2}+i\pi,\vartheta_{1})F_{2}^{\mathcal{O}_{2}}(\vartheta_{2}+i\pi,\vartheta_{1}')\\
 &  & +\left\{ \mathcal{O}^{1}\leftrightarrow\mathcal{O}^{2}\right\} \Big\}\left[1-S(\vartheta_{2}-\vartheta_{1})S(\vartheta_{1}'-\vartheta_{1})\right]\left[S(\vartheta_{1}-\vartheta_{1}')-S(\vartheta_{1}-\vartheta_{2})\right]\\
 & - & i\iiint\frac{\mathrm{d}\vartheta_{1}}{2\pi}\frac{\mathrm{d}\vartheta_{2}}{2\pi}\frac{\mathrm{d}\vartheta_{1}'}{2\pi}\frac{K_{t,x}^{\left(R\right)}(\vartheta_{1},\vartheta_{2}|\vartheta_{1},\vartheta_{2}')}{\left(\vartheta_{2}-\vartheta_{1}\right)}\Big\{ F_{2}^{\mathcal{O}_{1}}(\vartheta_{1}+i\pi,\vartheta_{1}')F_{2}^{\mathcal{O}_{2}}(\vartheta_{2}+i\pi,\vartheta_{1}')\\
 &  & +\left\{ \mathcal{O}^{1}\leftrightarrow\mathcal{O}^{2}\right\} \Big\}\left[S(\vartheta_{2}-\vartheta_{1})-S(\vartheta_{1}-\vartheta_{1}')\right]\left[1-S(\vartheta_{1}'-\vartheta_{1})S(\vartheta_{1}-\vartheta_{2})\right]\\
 & + & \iiint\frac{\mathrm{d}\vartheta_{1}}{2\pi}\frac{\mathrm{d}\vartheta_{2}}{2\pi}\frac{\mathrm{d}\vartheta_{1}'}{2\pi}F_{2}^{\mathcal{O}_{1}}(\vartheta_{2}+i\pi,\vartheta_{1}')F_{2}^{\mathcal{O}_{2}}(\vartheta_{2}+i\pi,\vartheta_{1}')K_{t,x}^{\left(R\right)}(\vartheta_{1},\vartheta_{2}|\vartheta_{1}',\vartheta_{1})\\
 &  & \times\left\{ \left[mx\cosh\vartheta_{1}-imt\sinh\vartheta_{1}\right]T\left(\vartheta_{2},\vartheta_{1},\vartheta_{1}'\right)T\left(\vartheta_{1}',\vartheta_{1},\vartheta_{2}\right)\right.\\
 &  & \left.+\left[S(\vartheta_{2}-\vartheta_{1})S(\vartheta_{1}-\vartheta_{1}')-S(\vartheta_{1}'-\vartheta_{1})S(\vartheta_{1}-\vartheta_{2})\right]\varphi\left(\vartheta_{1}-\vartheta_{1}'\right)\right\} 
\end{eqnarray*}

\subsection{Other terms}

The following two terms in (\ref{eq:d22final})

\begin{eqnarray*}
 & - & \iint\frac{\mathrm{d}\vartheta_{1}}{2\pi}\frac{\mathrm{d}\vartheta_{2}}{2\pi}F_{2}^{\mathcal{O}_{1}}(\vartheta_{2}+i\pi,\vartheta_{1})F_{2}^{\mathcal{O}_{2}}(\vartheta_{2}+i\pi,\vartheta_{1})K_{t,x}^{\left(R\right)}(\vartheta_{1},\vartheta_{2}|\vartheta_{1},\vartheta_{1})\\
 & - & \int\frac{\mathrm{d}\vartheta_{1}}{2\pi}\int\frac{\mathrm{d}\vartheta_{1}'}{2\pi}F_{2}^{\mathcal{O}_{1}}(\vartheta_{1}+i\pi,\vartheta_{1}')F_{2}^{\mathcal{O}_{2}}(\vartheta_{1}+i\pi,\vartheta_{1}')K_{t,x}^{\left(R\right)}(\vartheta_{1},\vartheta_{1}|\vartheta_{1},\vartheta_{1}')
\end{eqnarray*}
transform to each other under the symmetry (\ref{eq:d22symmetry}).

For the remaining terms in (\ref{eq:d22final}), after the transformation
we redefine $\vartheta_{2}\leftrightarrow\vartheta_{1}'$ to have
the same $K$ factor as before and get:

\begin{eqnarray*}
 &  & \iint\frac{\mathrm{d}\vartheta_{1}}{2\pi}\frac{\mathrm{d}\vartheta_{2}}{2\pi}\int\frac{\mathrm{d}\vartheta_{1}'}{2\pi}K_{t,x}^{\left(R\right)}(\vartheta_{1},\vartheta_{2}|\vartheta_{1}',\vartheta_{1})\\
 &  & \times\left(F_{4rc}^{\mathcal{O}_{1}}(\vartheta_{1}+i\pi,\vartheta_{1}'+i\pi|\vartheta_{2},\vartheta_{1})F_{2}^{\mathcal{O}_{2}}(\vartheta_{2}+i\pi,\vartheta_{1}')+\left\{ \mathcal{O}^{1}\leftrightarrow\mathcal{O}^{2}\right\} \right)\\
 & - & i\iiint\frac{\mathrm{d}\vartheta_{1}}{2\pi}\frac{\mathrm{d}\vartheta_{2}}{2\pi}\frac{\mathrm{d}\vartheta_{1}'}{2\pi}\frac{K_{t,x}^{\left(R\right)}(\vartheta_{1},\vartheta_{2}|\vartheta_{1}',\vartheta_{1})}{\left(\vartheta_{1}'-\vartheta_{1}\right)}\left[S(\vartheta_{1}-\vartheta_{1}')-S(\vartheta_{2}-\vartheta_{1})\right]\\
 &  & \times\Big\{ F_{2}^{\mathcal{O}_{1}}(\vartheta_{1}+i\pi,\vartheta_{2})F_{2}^{\mathcal{O}_{2}}(\vartheta_{2}+i\pi,\vartheta_{1}')+\left\{ \mathcal{O}_{1}\leftrightarrow\mathcal{O}_{2}\right\} \Big\}\\
 & - & i\iint\frac{\mathrm{d}\vartheta_{1}}{2\pi}\frac{\mathrm{d}\vartheta_{2}}{2\pi}\wp\int\frac{\mathrm{d}\vartheta_{1}'}{2\pi}\frac{K_{t,x}^{\left(R\right)}(\vartheta_{1},\vartheta_{2}|\vartheta_{1}',\vartheta_{1})}{\left(\vartheta_{1}'-\vartheta_{2}\right)}\left[S(\vartheta_{2}-\vartheta_{1})S(\vartheta_{1}-\vartheta_{1}')-1\right]\\
 &  & \times\Big\{ F_{2}^{\mathcal{O}_{1}}(i\pi,0)F_{2}^{\mathcal{O}_{2}}(\vartheta_{2}+i\pi,\vartheta_{1}')+\left\{ \mathcal{O}_{1}\leftrightarrow\mathcal{O}_{2}\right\} \Big\}\\
 & - & i\iint\frac{\mathrm{d}\vartheta_{1}}{2\pi}\frac{\mathrm{d}\vartheta_{2}}{2\pi}\wp\int\frac{\mathrm{d}\vartheta_{1}'}{2\pi}\frac{K_{t,x}^{\left(R\right)}(\vartheta_{1},\vartheta_{2}|\vartheta_{1}',\vartheta_{1})}{\left(\vartheta_{1}-\vartheta_{2}\right)}\left[S(\vartheta_{2}-\vartheta_{1})-S(\vartheta_{1}-\vartheta_{1}')\right]\\
 &  & \times\Big\{ F_{2}^{\mathcal{O}_{1}}(\vartheta_{2}+i\pi,\vartheta_{1}')F_{2}^{\mathcal{O}_{2}}(\vartheta_{1}'+i\pi,\vartheta_{1})+\left\{ \mathcal{O}_{1}\leftrightarrow\mathcal{O}_{2}\right\} \Big\}\\
 & - & \iiint\frac{\mathrm{d}\vartheta_{1}}{2\pi}\frac{\mathrm{d}\vartheta_{2}}{2\pi}\frac{\mathrm{d}\vartheta_{1}'}{2\pi}F_{2}^{\mathcal{O}_{1}}(\vartheta_{2}+i\pi,\vartheta_{1}')F_{2}^{\mathcal{O}_{2}}(\vartheta_{2}+i\pi,\vartheta_{1}')K_{t,x}^{\left(R\right)}(\vartheta_{1},\vartheta_{2}|\vartheta_{1}',\vartheta_{1})\times\\
 &  & \times\left\{ T\left(\vartheta_{2},\vartheta_{1},\vartheta_{1}'\right)\left[-mx\cosh\vartheta_{1}+im\left(R-t\right)\sinh\vartheta_{1}\right]+\varphi(\vartheta_{2}-\vartheta_{1})S(\vartheta_{2}-\vartheta_{1})S(\vartheta_{1}-\vartheta_{1}')\right\} 
\end{eqnarray*}

\subsection{Collecting the terms}

Using the results so far, the transformed $D_{22}$ is the following:
\begin{eqnarray*}
D_{22}^{trans} & = & \frac{1}{4}\iiiint\frac{\mathrm{d}\vartheta_{1}}{2\pi}\frac{\mathrm{d}\vartheta_{2}}{2\pi}\frac{\mathrm{d}\vartheta_{1}'}{2\pi}\frac{\mathrm{d}\vartheta_{2}'}{2\pi}K_{t,x}^{\left(R\right)}(\vartheta_{1},\vartheta_{2}|\vartheta_{1}'+i\varepsilon,\vartheta_{2}'+i\varepsilon)\\
 &  & F_{4}^{\mathcal{O}_{1}}(\vartheta_{2}+i\pi,\vartheta_{1}+i\pi,\vartheta_{1}'+i\varepsilon,\vartheta_{2}'+i\varepsilon)F_{4}^{\mathcal{O}_{2}}(\vartheta_{1}+i\pi,\vartheta_{2}+i\pi,\vartheta_{2}'+i\varepsilon,\vartheta_{1}'+i\varepsilon)\\
 & + & (F_{4rc}\mbox{ terms})+(\mbox{first order pole terms)}+(\mbox{second order pole terms)}\\
 & - & \iint\frac{\mathrm{d}\vartheta_{1}}{2\pi}\frac{\mathrm{d}\vartheta_{2}}{2\pi}F_{2}^{\mathcal{O}_{1}}(\vartheta_{2}+i\pi,\vartheta_{1})F_{2}^{\mathcal{O}_{2}}(\vartheta_{2}+i\pi,\vartheta_{1})K_{t,x}^{\left(R\right)}(\vartheta_{1},\vartheta_{2}|\vartheta_{1},\vartheta_{1})\\
 & - & \int\frac{\mathrm{d}\vartheta_{1}}{2\pi}\int\frac{\mathrm{d}\vartheta_{1}'}{2\pi}F_{2}^{\mathcal{O}_{1}}(\vartheta_{1}+i\pi,\vartheta_{1}')F_{2}^{\mathcal{O}_{2}}(\vartheta_{1}+i\pi,\vartheta_{1}')K_{t,x}^{\left(R\right)}(\vartheta_{1},\vartheta_{1}|\vartheta_{1},\vartheta_{1}')\\
 & + & D_{22}^{diag}
\end{eqnarray*}
where the $F_{4rc}$ terms are

\begin{eqnarray}
 & + & \iiint\frac{\mathrm{d}\vartheta_{1}}{2\pi}\frac{\mathrm{d}\vartheta_{2}}{2\pi}\frac{\mathrm{d}\vartheta_{1}'}{2\pi}K_{t,x}^{\left(R\right)}(\vartheta_{1},\vartheta_{2}|\vartheta_{1}',\vartheta_{1})\Big\{ F_{4rc}^{\mathcal{O}_{1}}(\vartheta_{1}+i\pi,\vartheta_{2}+i\pi|\vartheta_{1}',\vartheta_{1})F_{2}^{\mathcal{O}_{2}}(\vartheta_{2}+i\pi,\vartheta_{1}')\nonumber \\
 &  & +\left\{ \mathcal{O}^{1}\leftrightarrow\mathcal{O}^{2}\right\} \Big\}\nonumber \\
 & + & \iiint\frac{\mathrm{d}\vartheta_{1}}{2\pi}\frac{\mathrm{d}\vartheta_{2}}{2\pi}\frac{\mathrm{d}\vartheta_{1}'}{2\pi}K_{t,x}^{\left(R\right)}(\vartheta_{1},\vartheta_{2}|\vartheta_{1}',\vartheta_{1})\Big\{ F_{4rc}^{\mathcal{O}_{1}}(\vartheta_{1}+i\pi,\vartheta_{1}'+i\pi|\vartheta_{2},\vartheta_{1})F_{2}^{\mathcal{O}_{2}}(\vartheta_{2}+i\pi,\vartheta_{1}')\nonumber \\
 &  & +\left\{ \mathcal{O}^{1}\leftrightarrow\mathcal{O}^{2}\right\} \Big\}\nonumber \\
 & - & \iiint\frac{\mathrm{d}\vartheta_{1}}{2\pi}\frac{\mathrm{d}\vartheta_{2}}{2\pi}\frac{\mathrm{d}\vartheta_{1}'}{2\pi}K_{t,x}^{\left(R\right)}(\vartheta_{1},\vartheta_{2}|\vartheta_{1}',\vartheta_{1})\Big\{ F_{4rc}^{\mathcal{O}_{1}}(\vartheta_{2}+i\pi,\vartheta_{1}+i\pi|\vartheta_{1},\vartheta_{1}')F_{2}^{\mathcal{O}_{2}}(\vartheta_{2}+i\pi,\vartheta_{1}')\nonumber \\
 &  & +\left\{ \mathcal{O}^{1}\leftrightarrow\mathcal{O}^{2}\right\} \Big\}\label{eq:F4rcterms}
\end{eqnarray}
the first order pole terms are 
\begin{eqnarray}
 & - & i\iint\frac{\mathrm{d}\vartheta_{1}}{2\pi}\frac{\mathrm{d}\vartheta_{2}}{2\pi}\wp\int\frac{\mathrm{d}\vartheta_{1}'}{2\pi}\frac{K_{t,x}^{\left(R\right)}(\vartheta_{1},\vartheta_{2}|\vartheta_{1}',\vartheta_{1})}{\left(\vartheta_{2}-\vartheta_{1}'\right)}\Big\{ F_{2}^{\mathcal{O}_{1}}(i\pi,0)F_{2}^{\mathcal{O}_{2}}(\vartheta_{2}+i\pi,\vartheta_{1}')\nonumber \\
 &  & +\left\{ \mathcal{O}^{1}\leftrightarrow\mathcal{O}^{2}\right\} \Big\}\left[S(\vartheta_{1}'-\vartheta_{1})-S(\vartheta_{2}-\vartheta_{1})\right]\left[S(\vartheta_{1}-\vartheta_{2})-S(\vartheta_{1}-\vartheta_{1}')\right]\nonumber \\
 & - & i\iint\frac{\mathrm{d}\vartheta_{1}}{2\pi}\frac{\mathrm{d}\vartheta_{2}}{2\pi}\wp\int\frac{\mathrm{d}\vartheta_{1}'}{2\pi}\frac{K_{t,x}^{\left(R\right)}(\vartheta_{1},\vartheta_{2}|\vartheta_{1}',\vartheta_{1})}{\left(\vartheta_{1}-\vartheta_{1}'\right)}\Big\{ F_{2}^{\mathcal{O}_{1}}(\vartheta_{2}+i\pi,\vartheta_{1})F_{2}^{\mathcal{O}_{2}}(\vartheta_{2}+i\pi,\vartheta_{1}')\nonumber \\
 &  & +\left\{ \mathcal{O}^{1}\leftrightarrow\mathcal{O}^{2}\right\} \Big\}\left[1-S(\vartheta_{2}-\vartheta_{1})S(\vartheta_{1}'-\vartheta_{1})\right]\left[S(\vartheta_{1}-\vartheta_{1}')-S(\vartheta_{1}-\vartheta_{2})\right]\nonumber \\
 & - & i\iiint\frac{\mathrm{d}\vartheta_{1}}{2\pi}\frac{\mathrm{d}\vartheta_{2}}{2\pi}\frac{\mathrm{d}\vartheta_{1}'}{2\pi}\frac{K_{t,x}^{\left(R\right)}(\vartheta_{1},\vartheta_{2}|\vartheta_{1},\vartheta_{2}')}{\left(\vartheta_{2}-\vartheta_{1}\right)}\Big\{ F_{2}^{\mathcal{O}_{1}}(\vartheta_{1}+i\pi,\vartheta_{1}')F_{2}^{\mathcal{O}_{2}}(\vartheta_{2}+i\pi,\vartheta_{1}')\nonumber \\
 &  & +\left\{ \mathcal{O}^{1}\leftrightarrow\mathcal{O}^{2}\right\} \Big\}\left[S(\vartheta_{2}-\vartheta_{1})-S(\vartheta_{1}-\vartheta_{1}')\right]\left[1-S(\vartheta_{1}'-\vartheta_{1})S(\vartheta_{1}-\vartheta_{2})\right]\nonumber \\
 & - & i\iiint\frac{\mathrm{d}\vartheta_{1}}{2\pi}\frac{\mathrm{d}\vartheta_{2}}{2\pi}\frac{\mathrm{d}\vartheta_{1}'}{2\pi}\frac{K_{t,x}^{\left(R\right)}(\vartheta_{1},\vartheta_{2}|\vartheta_{1}',\vartheta_{1})}{\left(\vartheta_{1}'-\vartheta_{1}\right)}\left[S(\vartheta_{1}-\vartheta_{1}')-S(\vartheta_{2}-\vartheta_{1})\right]\nonumber \\
 &  & \times\Big\{ F_{2}^{\mathcal{O}_{1}}(\vartheta_{1}+i\pi,\vartheta_{2})F_{2}^{\mathcal{O}_{2}}(\vartheta_{2}+i\pi,\vartheta_{1}')+\left\{ \mathcal{O}_{1}\leftrightarrow\mathcal{O}_{2}\right\} \Big\}\nonumber \\
 & - & i\iint\frac{\mathrm{d}\vartheta_{1}}{2\pi}\frac{\mathrm{d}\vartheta_{2}}{2\pi}\wp\int\frac{\mathrm{d}\vartheta_{1}'}{2\pi}\frac{K_{t,x}^{\left(R\right)}(\vartheta_{1},\vartheta_{2}|\vartheta_{1}',\vartheta_{1})}{\left(\vartheta_{1}'-\vartheta_{2}\right)}\left[S(\vartheta_{2}-\vartheta_{1})S(\vartheta_{1}-\vartheta_{1}')-1\right]\nonumber \\
 &  & \times\Big\{ F_{2}^{\mathcal{O}_{1}}(i\pi,0)F_{2}^{\mathcal{O}_{2}}(\vartheta_{2}+i\pi,\vartheta_{1}')+\left\{ \mathcal{O}_{1}\leftrightarrow\mathcal{O}_{2}\right\} \Big\}\nonumber \\
 & - & i\iint\frac{\mathrm{d}\vartheta_{1}}{2\pi}\frac{\mathrm{d}\vartheta_{2}}{2\pi}\wp\int\frac{\mathrm{d}\vartheta_{1}'}{2\pi}\frac{K_{t,x}^{\left(R\right)}(\vartheta_{1},\vartheta_{2}|\vartheta_{1}',\vartheta_{1})}{\left(\vartheta_{1}-\vartheta_{2}\right)}\left[S(\vartheta_{2}-\vartheta_{1})-S(\vartheta_{1}-\vartheta_{1}')\right]\nonumber \\
 &  & \times\Big\{ F_{2}^{\mathcal{O}_{1}}(\vartheta_{2}+i\pi,\vartheta_{1}')F_{2}^{\mathcal{O}_{2}}(\vartheta_{1}'+i\pi,\vartheta_{1})+\left\{ \mathcal{O}_{1}\leftrightarrow\mathcal{O}_{2}\right\} \Big\}\label{eq:firstorderpoleterms}
\end{eqnarray}
while the second order pole terms are 
\begin{eqnarray}
 &  & +\iiint\frac{\mathrm{d}\vartheta_{1}}{2\pi}\frac{\mathrm{d}\vartheta_{2}}{2\pi}\frac{\mathrm{d}\vartheta_{1}'}{2\pi}F_{2}^{\mathcal{O}_{1}}(\vartheta_{2}+i\pi,\vartheta_{1}')F_{2}^{\mathcal{O}_{2}}(\vartheta_{2}+i\pi,\vartheta_{1}')K_{t,x}^{\left(R\right)}(\vartheta_{1},\vartheta_{2}|\vartheta_{1}',\vartheta_{1})\nonumber \\
 &  & \times\left\{ \left[mx\cosh\vartheta_{1}-imt\sinh\vartheta_{1}\right]T\left(\vartheta_{2},\vartheta_{1},\vartheta_{1}'\right)T\left(\vartheta_{1}',\vartheta_{1},\vartheta_{2}\right)\right.\nonumber \\
 &  & \left.+\left[S(\vartheta_{2}-\vartheta_{1})S(\vartheta_{1}-\vartheta_{1}')-S(\vartheta_{1}'-\vartheta_{1})S(\vartheta_{1}-\vartheta_{2})\right]\varphi\left(\vartheta_{1}-\vartheta_{1}'\right)\right\} \nonumber \\
 &  & -\iiint\frac{\mathrm{d}\vartheta_{1}}{2\pi}\frac{\mathrm{d}\vartheta_{2}}{2\pi}\frac{\mathrm{d}\vartheta_{1}'}{2\pi}F_{2}^{\mathcal{O}_{1}}(\vartheta_{2}+i\pi,\vartheta_{1}')F_{2}^{\mathcal{O}_{2}}(\vartheta_{2}+i\pi,\vartheta_{1}')K_{t,x}^{\left(R\right)}(\vartheta_{1},\vartheta_{2}|\vartheta_{1}',\vartheta_{1})\times\label{eq:secondorderpoleterms}\\
 &  & \times\left\{ T\left(\vartheta_{2},\vartheta_{1},\vartheta_{1}'\right)\left[mx\cosh\vartheta_{1}+im\left(R-t\right)\sinh\vartheta_{1}\right]-\varphi(\vartheta_{2}-\vartheta_{1})S(\vartheta_{2}-\vartheta_{1})S(\vartheta_{1}-\vartheta_{1}')\right\} \nonumber 
\end{eqnarray}

\subsubsection{$F_{4rc}$ terms}

The last two $F_{4rc}$ terms (\ref{eq:F4rcterms}) cancel due to

\[
F_{4rc}^{\mathcal{O}_{1,2}}(\vartheta_{1}+i\pi,\vartheta_{1}'+i\pi|\vartheta_{2},\vartheta_{1})-F_{4rc}^{\mathcal{O}_{1,2}}(\vartheta_{2}+i\pi,\vartheta_{1}+i\pi|\vartheta_{1},\vartheta_{1}')=0
\]
which can be easily proven from the definition (\ref{eq:F4rc_definition});
the remaining one is combined with the first order pole terms below
into the $F_{4ss}$ contributions in (\ref{eq:d22final}).

\subsubsection{First order pole terms}

The first order pole terms (\ref{eq:firstorderpoleterms}) can be
rearranged into the form

\begin{eqnarray*}
 & - & i\iint\frac{\mathrm{d}\vartheta_{1}}{2\pi}\frac{\mathrm{d}\vartheta_{2}}{2\pi}\wp\int\frac{\mathrm{d}\vartheta_{1}'}{2\pi}\frac{K_{t,x}^{\left(R\right)}(\vartheta_{1},\vartheta_{2}|\vartheta_{1}',\vartheta_{1})}{\left(\vartheta_{2}-\vartheta_{1}'\right)}\Big\{ F_{2}^{\mathcal{O}_{1}}(i\pi,0)F_{2}^{\mathcal{O}_{2}}(\vartheta_{2}+i\pi,\vartheta_{1}')\\
 &  & +\left\{ \mathcal{O}^{1}\leftrightarrow\mathcal{O}^{2}\right\} \Big\}\left[S(\vartheta_{1}'-\vartheta_{1})S(\vartheta_{1}-\vartheta_{2})-1\right]\\
 & - & i\iint\frac{\mathrm{d}\vartheta_{1}}{2\pi}\frac{\mathrm{d}\vartheta_{2}}{2\pi}\wp\int\frac{\mathrm{d}\vartheta_{1}'}{2\pi}\frac{K_{t,x}^{\left(R\right)}(\vartheta_{1},\vartheta_{2}|\vartheta_{1}',\vartheta_{1})}{\left(\vartheta_{1}-\vartheta_{1}'\right)}\Big\{ F_{2}^{\mathcal{O}_{1}}(\vartheta_{2}+i\pi,\vartheta_{1})F_{2}^{\mathcal{O}_{2}}(\vartheta_{2}+i\pi,\vartheta_{1}')\\
 &  & +\left\{ \mathcal{O}^{1}\leftrightarrow\mathcal{O}^{2}\right\} \Big\}\left[S(\vartheta_{1}'-\vartheta_{1})-S(\vartheta_{1}-\vartheta_{2})\right]\\
 & - & i\iiint\frac{\mathrm{d}\vartheta_{1}}{2\pi}\frac{\mathrm{d}\vartheta_{2}}{2\pi}\frac{\mathrm{d}\vartheta_{1}'}{2\pi}\frac{K_{t,x}^{\left(R\right)}(\vartheta_{1},\vartheta_{2}|\vartheta_{1},\vartheta_{1}')}{\left(\vartheta_{2}-\vartheta_{1}\right)}\Big\{ F_{2}^{\mathcal{O}_{1}}(\vartheta_{1}+i\pi,\vartheta_{1}')F_{2}^{\mathcal{O}_{2}}(\vartheta_{2}+i\pi,\vartheta_{1}')\\
 &  & +\left\{ \mathcal{O}^{1}\leftrightarrow\mathcal{O}^{2}\right\} \Big\}\left[S(\vartheta_{1}-\vartheta_{2})-S(\vartheta_{1}'-\vartheta_{1})\right]
\end{eqnarray*}
using 
\[
S(\vartheta)S(-\vartheta)=1
\]
These terms are combined with the $F_{4rc}$ terms into the $F_{4ss}$
contributions in (\ref{eq:d22final}).

\subsubsection{Second order pole terms}

The second order pole terms (\ref{eq:secondorderpoleterms}) have
a dependence on $R-t$ on the last line. One can make an integration
by parts to transform it into a $t$-dependence as on the second line.
Using

\begin{eqnarray*}
\frac{\partial K_{t,x}^{\left(R\right)}(\vartheta_{1},\vartheta_{2}|\vartheta_{1}',\vartheta_{1})}{\partial\vartheta_{1}} & = & -mR\sinh\left(\vartheta_{1}\right)K_{t,x}^{\left(R\right)}(\vartheta_{1},\vartheta_{2}|\vartheta_{1}',\vartheta_{1})
\end{eqnarray*}
we can make the following transformation in the integrand

\begin{eqnarray*}
 &  & -iT\left(\vartheta_{2},\vartheta_{1},\vartheta_{1}'\right)mR\sinh\vartheta_{1}K_{t,x}^{\left(R\right)}(\vartheta_{1},\vartheta_{2}|\vartheta_{1}',\vartheta_{1})=iT\left(\vartheta_{2},\vartheta_{1},\vartheta_{1}'\right)\frac{\partial K_{t,x}^{\left(R\right)}(\vartheta_{1},\vartheta_{2}|\vartheta_{1}',\vartheta_{1})}{\partial\vartheta_{1}}\\
 &  & \mbox{(partial integration)}\Rightarrow-iK_{t,x}^{\left(R\right)}(\vartheta_{1},\vartheta_{2}|\vartheta_{1}',\vartheta_{1})\frac{T\left(\vartheta_{2},\vartheta_{1},\vartheta_{1}'\right)}{\partial\vartheta_{1}}\\
 &  & =-K_{t,x}^{\left(R\right)}(\vartheta_{1},\vartheta_{2}|\vartheta_{1}',\vartheta_{1})S(\vartheta_{2}-\vartheta_{1})S(\vartheta_{1}-\vartheta_{1}')\left[\varphi\left(\vartheta_{1}-\vartheta_{1}'\right)-\varphi\left(\vartheta_{2}-\vartheta_{1}\right)\right]
\end{eqnarray*}
and so the contribution is transformed into 
\begin{align*}
 & +\iiint\frac{\mathrm{d}\vartheta_{1}}{2\pi}\frac{\mathrm{d}\vartheta_{2}}{2\pi}\frac{\mathrm{d}\vartheta_{1}'}{2\pi}F_{2}^{\mathcal{O}_{1}}(\vartheta_{2}+i\pi,\vartheta_{1}')F_{2}^{\mathcal{O}_{2}}(\vartheta_{2}+i\pi,\vartheta_{1}')K_{t,x}^{\left(R\right)}(\vartheta_{1},\vartheta_{2}|\vartheta_{1}',\vartheta_{1})\\
 & \times\left\{ \left[mx\cosh\vartheta_{1}-imt\sinh\vartheta_{1}\right]T\left(\vartheta_{1}',\vartheta_{1},\vartheta_{2}\right)-S(\vartheta_{1}'-\vartheta_{1})S(\vartheta_{1}-\vartheta_{2})\varphi\left(\vartheta_{1}-\vartheta_{1}'\right)\right\} 
\end{align*}
With this the second order pole terms (\ref{eq:secondorderpoleterms})
reduce to the form of the second order pole term in (\ref{eq:d22final}).

\subsection{End result}

After putting together every term, using the definition of $F_{4ss}$
(\ref{eq:F4ssdef}), and performing some simplifications one obtains
\[
D_{22}^{trans}=D_{22}
\]
which is exactly what was to be proven.

\bibliographystyle{utphys}
\bibliography{finiteTcorr}

\end{document}